\documentclass[11pt]{article}

\usepackage[final]{acl}
\usepackage{times}
\usepackage{latexsym}

\usepackage[T1]{fontenc}
\usepackage[utf8]{inputenc}
\usepackage{microtype}
\usepackage{inconsolata}
\usepackage{graphicx}
\usepackage{booktabs}
\usepackage{colortbl}
\usepackage{xspace}
\usepackage[most]{tcolorbox}
\usepackage{tcolorbox}
\usepackage{array}
\usepackage{amsfonts}
\usepackage{amsmath}

\usepackage[dvipsnames]{xcolor}
\usepackage[table]{xcolor}
\definecolor{lightgray}{gray}{0.95}
\usepackage{amssymb}
\usepackage{multirow}
\usepackage{multicol}
\usepackage{listings}
\usepackage{placeins}
\usepackage{subfigure}
\usepackage{graphicx}
\usepackage{enumitem}
\usepackage{hyperref}
\usepackage{tabularx}
\usepackage{mathrsfs}
\usepackage{pifont}
\usepackage{todonotes}

\newcommand{\ours}[1]{\textsc{CrowdNotes+}}
\newcommand{\ds}[1]{\textsc{HealthNotes}}
\newcommand{\judge}[1]{\textsc{HealthJudge}}
\newcommand{\greenup}{\textcolor{Green}{\uparrow}}
\newcommand{\reddown}{\textcolor{Red}{\downarrow}}
\newcommand{\grayeven}{\textcolor{Gray}{--}}
\newtcolorbox{prompt}{
  colback=gray!5,
  colframe=gray!35!black
}

\title{Beyond the Crowd: LLM-Augmented Community Notes\\ for Governing Health Misinformation}

\author{
\textbf{Jiaying Wu\textsuperscript{1}\thanks{Equal Contribution}},
\textbf{Zihang Fu\textsuperscript{1}\footnotemark[1]},
\textbf{Haonan Wang\textsuperscript{1}},
\textbf{Fanxiao Li\textsuperscript{2}},
\\
\textbf{Jiafeng Guo\textsuperscript{3,4}},
\textbf{Preslav Nakov\textsuperscript{5}},
\textbf{Min-Yen Kan\textsuperscript{1}}\\[3pt]
\textsuperscript{1}National University of Singapore, \textsuperscript{2}Yunnan University, \\
\textsuperscript{3}State Key Laboratory of AI Safety, Institute of Computing Technology, \\ Chinese Academy of Sciences, 
\textsuperscript{4}University of Chinese Academy of Sciences, \\
\textsuperscript{5}Mohamed bin Zayed University of Artificial Intelligence \\[3pt]
\texttt{jiayingwu@u.nus.edu, zihangfu@u.nus.edu, kanmy@comp.nus.edu.sg} 
  }
  
\begin{document}
\maketitle

\begin{abstract}
Community Notes, the crowd-sourced misinformation governance system on X (formerly Twitter), allows users to flag misleading posts, attach contextual notes, and rate the notes' helpfulness. However, our empirical analysis of 30.8K health-related notes reveals \textbf{substantial latency}, with a median delay of 17.6 hours before notes receive a helpfulness status. To improve responsiveness during real-world misinformation surges, we propose \ours{}, a unified LLM-based framework that augments Community Notes for faster and more reliable health misinformation governance. \ours{} integrates two modes: \textbf{(1)} evidence-grounded note \textbf{augmentation} and \textbf{(2)} utility-guided note \textbf{automation}, supported by a hierarchical three-stage evaluation of relevance, correctness, and helpfulness. We instantiate the framework with \ds{}, a benchmark of 1.2K health notes annotated for helpfulness, and a fine-tuned helpfulness judge. Our analysis first \textbf{uncovers a key loophole} in current crowd-sourced governance: voters frequently conflate stylistic fluency with factual accuracy. Addressing this via our hierarchical evaluation, experiments across 15 representative LLMs demonstrate that \ours{} significantly outperforms human contributors in \textbf{note correctness, helpfulness, and evidence utility}.\footnote{Code and data are available at: \url{https://github.com/jiayingwu19/CrowdNotesPlus}.}
\end{abstract}

\section{Introduction}

\begin{figure}[t]
    \centering
    \includegraphics[width=0.9\columnwidth]{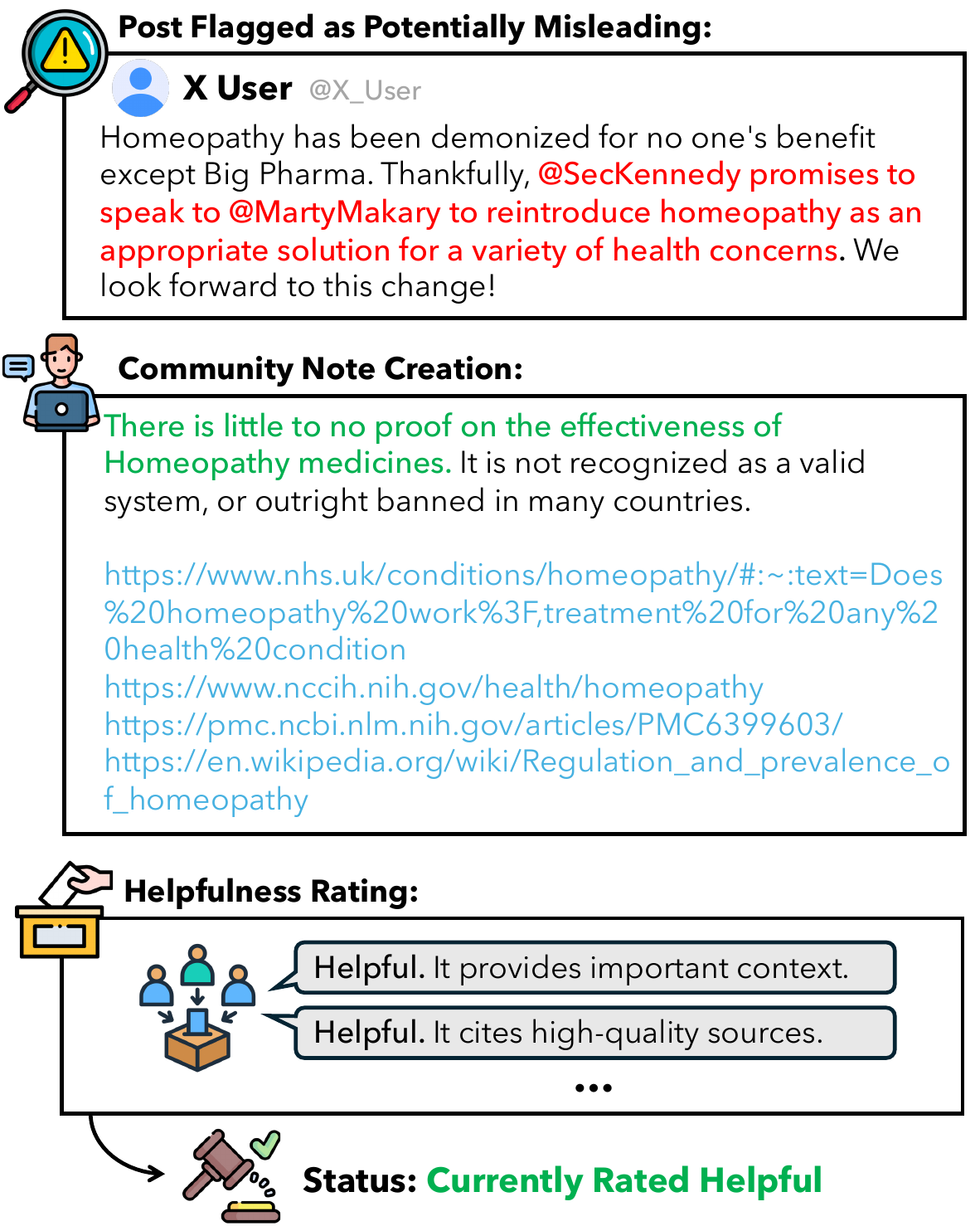}
\caption{\textbf{Overview of Community Notes on X for crowd-sourced misinformation governance.} Users engage in three stages: \textbf{(1) flagging} potentially misleading posts, \textbf{(2) writing} notes that provide clarification or additional context, and \textbf{(3) rating} the notes' helpfulness. Based on accumulated ratings, each note receives one of three statuses: (a)~\textit{Needs More Ratings}, (b)~\textit{Currently Rated Not Helpful}, or (c)~\textit{Currently Rated Helpful}. Only notes from the last category are publicly displayed alongside the original post to inform readers.}
\label{fig:intro}
\end{figure}

Health misinformation on social media has fueled persistent ``infodemics'' that undermine public trust and threaten individual well-being \cite{islam2020covid,shahbazi2024social}. Often triggered by major real-world events \cite{shahi2021exploratory,adebesin2023role}, such misinformation propagates at a scale and speed that outpace expert fact-checking and platform moderation \cite{godel2021moderating,singer2023closing}.

In response, \textbf{crowd-sourced fact-checking}, which leverages the collective wisdom of online contributors \cite{allen2021scaling,martel2024crowds,shahbazi2024social,pfander2025spotting}, has emerged as a scalable complement to expert-driven approaches.
Community Notes \cite{wojcik2022birdwatch} on X (formerly Twitter) is the most prominent example (Figure~\ref{fig:intro}). It allows users to flag posts, write contextual notes, and vote on helpfulness; only notes rated \textit{Currently Rated Helpful} are shown to the public.

While prior work has demonstrated Community Notes’ potential for improving discourse quality and reducing polarization \cite{chuai2024community,renault2024collaboratively,slaughter2025community}, our large-scale analysis of 30.8K health-related notes over four years (\S \ref{sec:cn-landscape}) reveals two systemic bottlenecks that limit the system’s responsiveness to fast-moving health misinformation: \textbf{(1) Delayed note generation.} Extending earlier reports of latency in Community Notes \cite{renault2024collaboratively}, we find that the first note appears a median of 10.4 hours after a misleading health post is flagged, and the first helpfulness verdict (i.e., \textit{Helpful/Not Helpful}) arrives another 7.2 hours later—well past the period of peak public attention.  \textbf{(2) Sparse helpfulness evaluation.} A striking 87.9\% of health notes remain indefinitely in the \textit{Needs More Ratings} state. As only \textit{Helpful} notes are surfaced, this bottleneck further delays corrective information from reaching users when it is most needed.

To address these limitations, we introduce \ours{}, a unified framework that leverages large language models (LLMs) to enhance both the creation and evaluation of Community Notes for more timely and reliable misinformation governance. Given a flagged post containing a potentially misleading claim, \ours{} extends the existing crowd-sourced pipeline through two complementary generation modes (Figure~\ref{fig:overview}): \textbf{(1) Evidence-Grounded Note Augmentation}, where humans supply evidence (e.g., URLs) and LLMs synthesize it into structured notes, and \textbf{(2) Utility-Guided Note Automation}, where LLMs autonomously plan, retrieve, and select high-quality evidence before generating notes. To ensure robust and interpretable assessment, \ours{} further incorporates a \textbf{hierarchical three-step evaluation pipeline} that progressively verifies \textbf{(1)} the \textit{relevance} of the retrieved evidence, \textbf{(2)} the \textit{correctness} of the evidence presented, and \textbf{(3)} the overall \textit{helpfulness} of the generated note.

We instantiate \ours{} in the health domain with \ds{}, a benchmark of 1.2K health-related Community Notes labeled \textit{Helpful} or \textit{Not Helpful}, along with \judge{}, a fine-tuned helpfulness evaluator. Experiments on fifteen LLMs validate the framework’s reliability and utility. We also identify a key weakness in crowd-sourced evaluation (\S \ref{sec:eval-superiority}), where stylistic fluency is often mistaken for factual accuracy, and show that our hierarchical evaluation reduces such false positives.

Across both generation modes, LLMs produce notes that are more accurate and better grounded than human-written notes (\S \ref{sec:gen-effectiveness}), while utility-guided automation consistently selects higher-quality evidence than human contributors (\S \ref{sec:evidence-utility}). Together, these improvements enhance both note reliability and evidential support. These results position \ours{} as a principled approach for improving the timeliness, factual consistency, and interpretability of crowd-sourced health misinformation governance on social media.

\section{Related Work}

\textbf{Crowd-Sourced Fact-Checking.}
The scale and speed of online misinformation make it unrealistic to rely solely on professional fact-checkers \cite{godel2021moderating,singer2023closing}. Crowd-sourced fact-checking \cite{allen2021scaling,martel2024crowds,shahbazi2024social,pfander2025spotting,xing2026communitynotes}, exemplified by Community Notes on X, allows users to collaboratively provide clarifications on potentially misleading content. Prior work shows that such community moderation can reduce misinformation engagement \cite{chuai2024did,slaughter2025community} and promote more balanced discourse \cite{chuai2024community,renault2024collaboratively}. However, most studies assume that notes already exist and focus on voting dynamics, consensus formation, or downstream impact. The earlier stage of note creation, especially in time-sensitive contexts, remains underexplored. Initial automation attempts \cite{de2025supernotes,singh2025on} have limited practicality because \cite{de2025supernotes} requires multiple human-written notes for the same post, and \cite{singh2025on} depends solely on LLM internal knowledge without web access, insufficient for emerging or unseen claims. Our work fills this gap in the health domain, where timeliness is crucial, by introducing a unified framework for systematic LLM-augmented note generation and evaluation.

\begin{figure*}[t]
    \centering
    \includegraphics[width=0.95\textwidth]{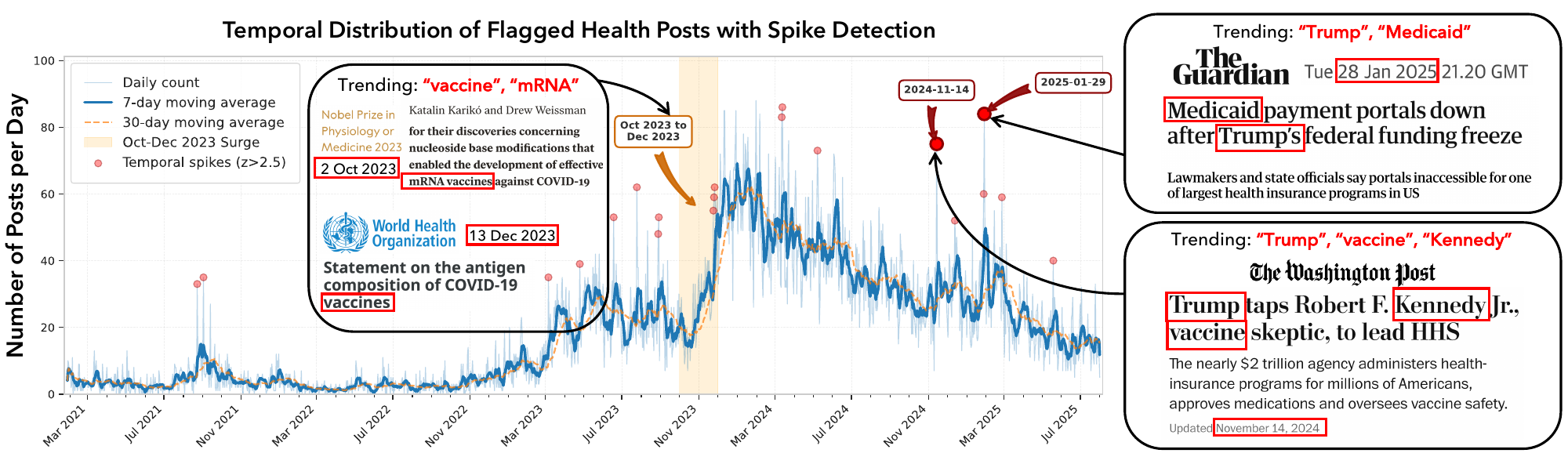}
\caption{\textbf{Spikes in flagged health misinformation posts align with major real-world health events} (detailed in \S\ref{sec:misinfo-dynamics}), including outbreak alerts, vaccine developments, and policy debates, highlighting the event-driven nature of misinformation activity on social media.}
    \label{fig:temporal-trend}
\end{figure*}
\textbf{Automated Governance of Textual Misinformation.}
Automated approaches aim to identify and counter misinformation at scale. Prior work has developed classifiers for detecting misleading posts and articles, using linguistic features \cite{potthast2018stylometric,zhang2021mining} and network-based signals \cite{wu2023decor,wu2023prompt}. While effective for flagging suspicious content, these systems rarely provide explanations that clarify why the content is misleading. 

Recent studies use LLMs to generate explanatory text \cite{hu2024bad,wu2024fake} and retrieve evidence from credible sources to justify predictions \cite{pan2023fact,zhang2023towards,zhou2024correcting}. However, these methods typically position the model as an autonomous arbiter, treating explanations merely as justifications. This overlooks the ``human-in-the-loop'' nature of governance systems like Community Notes. Our work bridges this gap by evaluating LLMs not as replacements, but as assistants that empower contributors with evidence-grounded drafts, preserving the human locus of control.

\section{Temporal Dynamics of Health Misinformation and Community Notes}

\label{sec:cn-landscape}

Understanding how health misinformation emerges and how community governance responds is essential for designing timely interventions. Before developing automated support, we analyze the temporal dynamics of health-related Community Notes on X to identify when misinformation surges occur and how promptly the system reacts.

\vspace{-5pt}
\subsection{Data Scope}
\label{sec:data-scope}
We collected all publicly available, user-contributed Community Notes\footnote{\url{https://x.com/i/communitynotes/download-data}} on X up to 4 August 2025, retaining only English entries for consistency. To focus on health-related content, we define seven topical categories: \textbf{(1)} diseases or medical conditions, \textbf{(2)} drugs, vaccines, treatments, and tests, \textbf{(3)} public health guidance or policy, \textbf{(4)} wellness products, diets, and supplements, \textbf{(5)} healthcare professionals or systems, \textbf{(6)} biological or epidemiological concepts, and \textbf{(7)} health-related conspiracies or hoaxes.

We filter the collected notes using zero-shot prompting with Lingshu-32B \cite{li2025scaling}, a multimodal LLM with state-of-the-art performance on medical QA. To validate this filter, we cross-check its predictions against closed-source LLMs on a random sample of 1,000 notes, observing high agreement (GPT-4.1 \cite{openai2025gpt4_1}: 99.2\%, Gemini-2.5-Flash \cite{google2025gemini_2_5}: 100\%, Claude-4-Sonnet \cite{anthropic2025claude4}: 96.8\%). Given this high reliability, we retain all notes classified as health-related by Lingshu-32B. We then retrieve the associated posts, using GPT-4.1 to keep only those with text-based health claims, while removing unavailable posts or URL-only content.
 
This process yields 30,791 health-related notes covering 25,484 potentially misleading posts.  We base our following analysis of temporal trends and systemic bottlenecks on this data.

\vspace{-5pt}
\subsection{Event-Driven Misinformation Dynamics}
\label{sec:misinfo-dynamics}

\begin{figure*}[t]
    \centering
    \includegraphics[width=\textwidth]{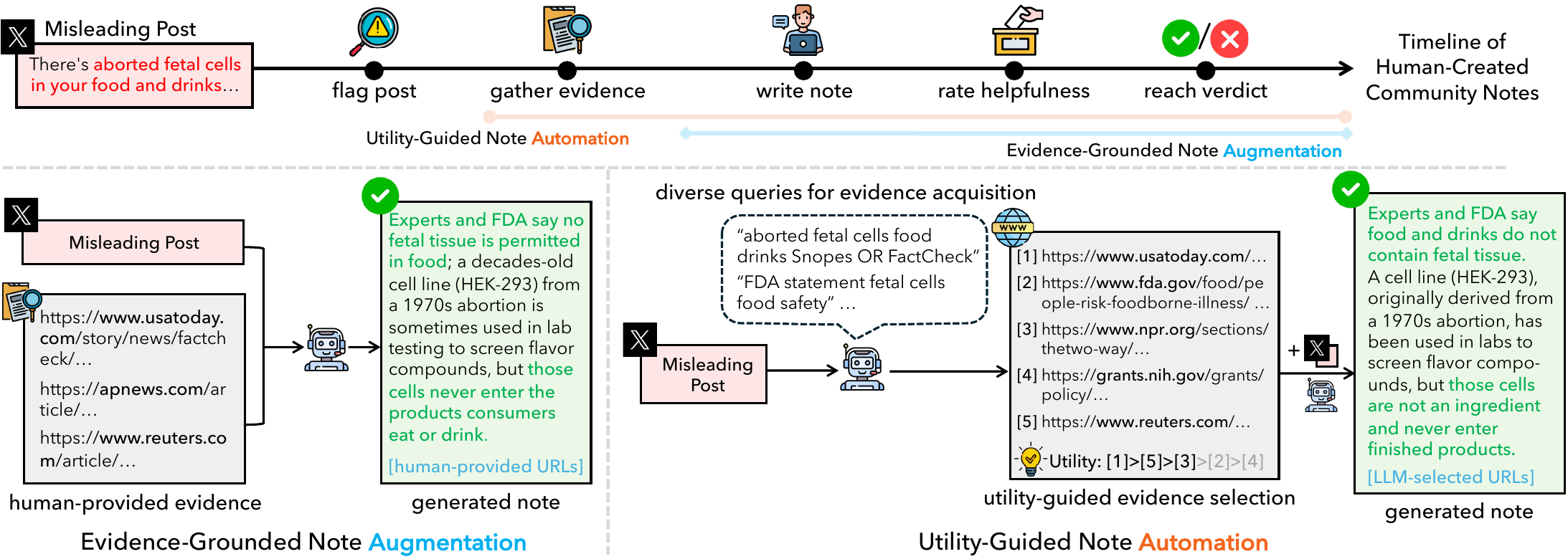}
\caption{\textbf{Overview of the proposed \ours{} framework for LLM-augmented Community Notes.} The upper timeline illustrates the human-created Community Notes workflow on X. The lower panels depict two note generation modes in \ours{}: \textbf{(1) evidence-grounded note augmentation}, where LLMs generate notes from human-provided evidence, and \textbf{(2) utility-guided note automation}, where LLMs autonomously retrieve and select high-utility evidence from the Web to generate notes.}    
\label{fig:overview}
\end{figure*}

We first examine the temporal distribution of the 25K health-related flagged posts to understand \textbf{how activity evolves relative to real-world events}. Daily post counts are compared against a 28-day rolling baseline, and a day is marked as a \textbf{spike} if its count exceeds the rolling mean by more than 2.5 standard deviations.

To contextualize each spike, we identify trending topics within a three-day window centered on the spike. We compute word frequencies from post text after removing stopwords, identify trending terms, and use these terms to characterize the dominant themes, associating each surge with major health events reported by mainstream news outlets or public health authorities during the same period. Only events that are uniquely prominent within their window are retained to avoid cross-period overlap and temporal ambiguity.

As illustrated by the spikes on 14 November 2024 and 29 January 2025, and the sustained rise from October to December 2023 (Figure~\ref{fig:temporal-trend}), \textit{misinformation activity aligns closely with major health developments}, including outbreak announcements, vaccine policy updates, and high-profile public health debates. These patterns show that health misinformation is strongly event-driven and emerges rapidly in response to external developments, motivating our next analysis on how quickly Community Notes respond once such posts appear.

\vspace{-5pt}
\subsection{Delays in Note Creation and Visibility}
\label{sec:note-delays}

\begin{table}[t]
\centering
\resizebox{\columnwidth}{!}{%
\begin{tabular}{lcc}
\toprule
\textbf{Pct.} & \textbf{Post Published $\rightarrow$ First Note} & \textbf{First Note $\rightarrow$ First Verdict} \\
\midrule
25\% & 3.4 & 3.6 \\
50\% & 10.4 & 7.2 \\
75\% & 23.0 & 18.4 \\
90\% & 49.1 & 76.4 \\
\bottomrule
\end{tabular}
}
\caption{\textbf{Delays (hours) in health Community Notes}, with a median of 17.6 hours before the first note attains any helpfulness verdict (i.e., \emph{Helpful} vs. \emph{Not Helpful}).}
\label{tab:time-delays}
\end{table}

Building on this analysis, we examine the 30K associated health-related Community Notes to assess \textbf{how quickly corrective information becomes visible}. Although Community Notes are intended to support timely, crowd-sourced fact-checking, our temporal analysis shows substantial delays. As reported in Table~\ref{tab:time-delays}, the median time from a misleading post to the creation of the first note is 10.4 hours. The subsequent voting phase adds another 7.2 hours before the note receives a helpfulness verdict (\textit{Helpful} or \textit{Not Helpful}). 

Furthermore, 87.9 percent of notes actually never gather enough votes to exit \textit{Needs More Ratings}, which prevents them from attaining any public-facing status.

Since only notes achieving \textit{Helpful} status are eventually surfaced to readers, \textit{these delays significantly restrict the availability of corrective information at critical moments when misinformation is spreading most rapidly and widely, limiting timely user awareness}. Improving responsiveness therefore requires accelerating both note creation and note evaluation while preserving factual rigor and consistency. This motivates our proposed framework, \ours{}, which leverages LLMs to enhance the timeliness, reliability, and scalability of Community Notes in dynamic, high-volume misinformation settings.

\section{\ours{}: Framework for LLM-Augmented Community Notes}
\label{sec:framework}

Our analysis shows that although health misinformation closely follows real-world events, the Community Notes workflow often lags behind due to slow note creation and delayed voting. To address these, we propose \ours{}, a unified framework that uses LLMs to accelerate note creation and evaluation.  \ours{} supports two complementary modes (Figure~\ref{fig:overview}): \textbf{(1)} evidence-grounded note augmentation and \textbf{(2)} utility-guided note automation, together with a hierarchical evaluation pipeline that assesses relevance, correctness, and helpfulness.

\subsection{Evidence-Grounded Note Augmentation}
\label{sec:llm-augmentation}
We first examine \textit{whether LLMs can assist contributors in the standard Community Notes setting where reliable evidence is manually provided.} In this workflow, a user flags a potentially misleading post $p$ and supplies a set of sources $\mathcal{E}_h$, where each $e \in \mathcal{E}_h$ is a URL linking to external content.

Each evidence piece $e$ is processed through a $\mathsf{RETRIEVE}$ step that segments its textual content into passages. Using the post $p$ as a query, a $\mathsf{MATCH}$ step selects the most relevant passage from each source, producing a set of evidence chunks $\mathcal{C}_h$. The model then executes a $\mathsf{GENERATE}$ step, conditioning on both $p$ and $\mathcal{C}_h$ to synthesize a concise, informativs note $n_h$. The evidence URLs $\mathcal{E}_h$ are attached after $n_h$ for transparency.

Figure \ref{fig:aug-example} presents a concrete example of this mode. It preserves the factual grounding of human-curated sources while automating the synthesis of concise, well-structured notes, reducing the time and effort required for human-written explanations. 

\vspace{-5pt}
\subsection{Utility-Guided Note Automation}
\label{sec:llm-automation}

We next examine \textit{whether note creation can be fully automated once a post $p$ is flagged as potentially misleading}, simulating a practical deployment scenario. Unlike the augmentation mode (\S \ref{sec:llm-augmentation}), this mode requires the model to retrieve, select, and synthesize evidence without human guidance.

Motivated by findings that diverse query formulations yield complementary retrieval results \cite{santos2015search,wu2024result}, the model generates a set of semantically diverse search queries $\mathcal{Q}$ from $p$. Each query retrieves top-ranked documents through a $\mathsf{SEARCH}$ step, and all retrieved items are merged and de-duplicated into a candidate pool $\mathcal{P} = \text{dedup}\left(\bigcup_{q \in \mathcal{Q}} \text{TopK}(q)\right)$.

To select informative evidence, we add an LLM-based \textbf{utility judgment} module inspired by evidence ranking \cite{zhang2024large}. Given a quota $\tau$, it iteratively selects and removes the highest-utility evidence snippets (title and summary), forming the set $\mathcal{E}_m$, whose URLs are appended for transparency. We then apply $\mathsf{RETRIEVE}$ and $\mathsf{MATCH}$ (\S \ref{sec:llm-augmentation}) to obtain chunks $\mathcal{C}_m$ and generate a note $n_m$ conditioned on $p$ and $\mathcal{C}_m$.

Figure \ref{fig:auto-example} illustrates the full pipeline and evidence selection behavior. This end-to-end mode enables fully automated note generation guided by evidence utility, reducing reliance on human effort while maintaining factual grounding. 

\subsection{Hierarchical Helpfulness Evaluation}
\label{sec:hierarchical-eval}

To ensure robust and interpretable assessment of the generated notes, \ours{} employs a three-stage evaluation pipeline that sequentially verifies \textbf{(1)} relevance, \textbf{(2)} correctness, and \textbf{(3)} helpfulness. 

\textbf{Relevance} evaluates \textit{whether the retrieved evidence offers meaningful factual context, clarification, or supporting information} that helps readers better assess the claim made in the post. It forms the foundation of retrieval-augmented generation \cite{saad2024ares,yu2025evaluation}, ensuring that notes are grounded in appropriate information.

\textbf{Correctness} evaluates \textit{whether the note faithfully represents the content of the cited sources}, without factual errors, exaggeration, or selective framing. Even when evidence is relevant, its interpretation can still be distorted, a common issue in scientific and medical communication \cite{glockner2024missci,wuehrl2024understanding}. This step ensures that the note’s claims align with the provided sources rather than relying on misinterpretation.

\textbf{Helpfulness} evaluates \textit{whether the note assists readers in understanding or critically evaluating the flagged post}, following the official Community Notes criteria.\footnote{\url{https://communitynotes.x.com/guide/en/under-the-hood/download-data}}

\paragraph{Operationalizing the Hierarchy.} We implement these criteria as sequential binary gates using LLM-based judges (implementation details in \S\ref{sec:benchmark} and Appendix \ref{app:eval-detail}). A note is evaluated for correctness only if it is deemed relevant, and for helpfulness only if it is correct. Formally, let $R$, $C$, and $H$ denote binary indicators of relevance, correctness, and helpfulness. The joint probability of a note satisfying all criteria decomposes as:

\vspace{-10pt}
{\small
\begin{align}
P(R{=}1, C{=}1, H{=}1)
&= P(H{=}1 \mid C{=}1, R{=}1) \nonumber \\
&\quad \times P(C{=}1 \mid R{=}1) \nonumber \\
&\quad \times P(R{=}1).
\end{align}
}

This formulation enforces a strict dependency: \textbf{a note is deemed helpful only if strictly grounded in relevance and correctness.} By decomposing helpfulness into these conditional components, our design prevents the common failure mode where models rely on surface-level fluency rather than factual reasoning \cite{wan2025truth}, yielding a transparent and fine-grained assessment.

\section{The \ds{} Benchmark}
\label{sec:benchmark}

We introduce \ds{}, the first benchmark for studying LLM-augmented Community Notes in the health domain. \ds{} combines a curated dataset with a customized evaluation judge, providing a reproducible foundation for analyzing LLM augmentation and automation methods in this high-stakes setting.

\textbf{Data.} To capture both successful and unsuccessful corrections, we include both \textit{Helpful} and \textit{Not Helpful} health notes as labeled by human contributors. From the health-related Community Notes collected in \S \ref{sec:data-scope}, we identify 3,713 notes with crowd-confirmed helpfulness labels (\textit{Helpful}: 2,971; \textit{Not Helpful}: 742). Among these, 634 \textit{Not Helpful} notes retain valid evidence URLs. To create a balanced benchmark, we sample an equal number of \textit{Helpful} notes, resulting in 1,268 post–note pairs.

Each data instance contains a flagged post, a corresponding note text, and verified evidence URLs. Table~\ref{tab:data-stats} summarizes the dataset statistics such as number of posts and evidence snippets. Figure~\ref{fig:topic-dist} shows the distribution over the seven health categories defined in \S \ref{sec:data-scope}, confirming that \ds{} covers diverse health-related topics (See Appendix \ref{app:healthnotes}).

\textbf{Evaluation Pipeline.}
Our evaluation follows the hierarchical scheme in \S\ref{sec:hierarchical-eval}. For \textit{relevance} and \textit{correctness}, we use an LLM-as-a-Judge setup with GPT-4.1 \cite{openai2025gpt4_1}. For the final \textit{helpfulness} stage, we introduce \judge{}, a fine-tuned Lingshu-7B model \cite{li2025scaling} designed for domain-specific note helpfulness assessment. We provide training details, human validation of judge reliability, and comparative performance results on helpfulness judgment in Appendix \ref{app:judge-reliability}.

\begin{table*}[t]
\centering
\resizebox{\textwidth}{!}{%
\begin{tabular}{ll
                c>{\columncolor{lightgray}}c
                c c>{\columncolor{lightgray}}c|
                c>{\columncolor{lightgray}}c
                c c>{\columncolor{lightgray}}c|
                cc}
\toprule
\multirow{3}{*}{} &  &
\multicolumn{5}{c|}{\textbf{Helpful (634)}} &
\multicolumn{5}{c|}{\textbf{Not Helpful (634)}} &
\multicolumn{2}{c}{\textbf{Overall}} \\   
\cmidrule(lr){3-7} \cmidrule(lr){8-12} \cmidrule(lr){13-14}
& \textbf{Setting} $\rightarrow$& \multicolumn{2}{c}{\textbf{Note Aug. (R=89.27)}} & \multicolumn{3}{c|}{\textbf{Note Auto.}}
  & \multicolumn{2}{c}{\textbf{Note Aug. (R=71.45)}} & \multicolumn{3}{c|}{\textbf{Note Auto.}} &
  \textbf{Aug.} & \textbf{Auto.} \\ 
\cmidrule(lr){3-4} \cmidrule(lr){5-7} \cmidrule(lr){8-9} \cmidrule(lr){10-12} \cmidrule(lr){13-13} \cmidrule(lr){14-14}
& \textbf{Model} $\downarrow$ & \textbf{C} & \textbf{H} & \textbf{R} & \textbf{C} & \textbf{H}
  & \textbf{C} & \textbf{H} & \textbf{R} & \textbf{C} & \textbf{H} & \textbf{H} & \textbf{H} \\   
\midrule
\multirow{1}{*}{} 
& Human Baseline         & 75.24 & 73.19 & 89.27 & 75.24 & 73.19   & 44.32 & 5.52 & 71.45 & 44.32 & 5.52   & \multicolumn{2}{c}{39.36} \\
\midrule
\multirow{3}{*}{\textbf{G1}} 
& Gemini-2.5-pro\dag     & \textbf{88.64} & 85.65 $\greenup$ & \textbf{95.74} & \underline{93.85} & \underline{91.17} $\greenup$ & \textbf{70.50} & 37.54 $\greenup$ & 91.96 & \underline{90.22} & 69.24 $\greenup$ & 61.60 $\greenup$ & \underline{80.21} $\greenup$ \\
& o3\dag               & 87.70 & \textbf{86.91} $\greenup$ & \textbf{95.74} & \textbf{94.16} & \textbf{92.11} $\greenup$& 68.30 & \textbf{40.69} $\greenup$ & 91.96 & 89.91 & \textbf{70.19} $\greenup$& \textbf{63.80} $\greenup$& \textbf{81.15} $\greenup$\\
& Grok-4\dag           & 86.44 & 82.65 $\greenup$ & \textbf{95.74} & 92.74 & 88.17 $\greenup$& 67.98 & 32.81 $\greenup$ & 91.96 & 89.27 & 67.19 $\greenup$& 57.73 $\greenup$& 77.68 $\greenup$\\
\midrule
\multirow{2}{*}{\textbf{G2}} 
& GPT-4.1               & \underline{87.85} & \underline{85.80} $\greenup$ & \underline{94.64} & 92.90 & 88.49 $\greenup$& \underline{69.56} & \underline{40.22} $\greenup$ & \underline{93.06} & \textbf{90.85} & \underline{69.87} $\greenup$& \underline{63.01} $\greenup$& 79.18 $\greenup$\\
& Claude-4-Opus         & 85.17 & 83.60 $\greenup$ & \underline{94.64} & 89.43 & 85.96 $\greenup$& 63.88 & 37.85 $\greenup$ & \underline{93.06} & 84.70 & 64.51 $\greenup$& 60.73 $\greenup$& 75.24 $\greenup$\\
\midrule
\multirow{6}{*}{\textbf{G3}} 
& Qwen3-32B           & 81.39 & 76.66 $\greenup$ & 90.69 & 80.28 & 70.35 $\reddown$& 60.57 & 28.86 $\greenup$ & 87.22 & 77.13 & 55.84 $\greenup$& 52.76 $\greenup$& 63.10 $\greenup$\\
& Qwen3-14B             & 76.03 & 70.82 $\reddown$ & 90.69 & 76.03 & 66.09 $\reddown$& 56.15 & 23.03 $\greenup$ & 87.22 & 71.29 & 50.63 $\greenup$& 46.93 $\greenup$& 58.36 $\greenup$\\
& Llama-3.1-8B        & 67.98 & 61.36 $\reddown$ & 86.59 & 60.41 & 49.05 $\reddown$& 51.10 & 17.98 $\greenup$ & 83.75 & 61.83 & 36.28 $\greenup$& 39.67 $\greenup$& 42.67 $\greenup$\\
& Ministral-8B           & 56.94 & 51.58 $\reddown$ & 86.59 & 53.31 & 44.32 $\reddown$& 43.22 & 14.67 $\greenup$ & 83.75 & 51.74 & 27.60 $\greenup$& 33.13 $\reddown$& 35.96 $\reddown$\\
& Qwen3-8B\dag          & 70.35 & 64.67 $\reddown$ & 86.59 & 65.30 & 53.63 $\reddown$& 47.00 & 18.14 $\greenup$ & 83.75 & 58.83 & 34.86 $\greenup$& 41.41 $\greenup$& 44.25 $\greenup$\\
& Qwen3-8B           & 69.56 & 64.83 $\reddown$ & 86.59  & 65.62 & 55.36 $\reddown$& 47.63 & 19.09 $\greenup$ & 83.75  & 61.20 & 38.80 $\greenup$& 41.96 $\greenup$& 47.08 $\greenup$\\
\midrule
\multirow{4}{*}{\textbf{G4}} 
& Lingshu-32B           & 79.34 & 73.19 \grayeven & 91.96 & 78.70 & 67.35 $\reddown$& 58.99 & 22.08 $\greenup$ & \textbf{93.85} & 81.70 & 52.37 $\greenup$& 47.64 $\greenup$& 59.86 $\greenup$\\
& MedGemma-27B           & 84.38 & 79.02 $\greenup$ & 91.96  & 85.96 & 79.81 $\greenup$& 65.46 & 30.91 $\greenup$ & \textbf{93.85} & 86.91 & 58.68 $\greenup$& 54.97 $\greenup$& 69.25 $\greenup$\\
& Lingshu-7B           & 58.04 & 50.47 $\reddown$ & 85.65   & 53.63 & 41.80 $\reddown$& 43.38 & 13.56 $\greenup$ & 85.33  & 60.41 & 33.91 $\greenup$& 32.02 $\reddown$& 37.86 $\reddown$\\
& MedGemma-4B           & 60.41 & 52.68 $\reddown$ & 85.65 & 53.63 & 40.06 $\reddown$& 43.53 & 16.56 $\greenup$ & 85.33& 56.31 & 31.23 $\greenup$& 34.62 $\reddown$& 35.65 $\reddown$\\
\bottomrule
\end{tabular}}
\caption{\textbf{Effectiveness (\%) of 15 representative LLMs across note augmentation (\S \ref{sec:llm-augmentation}) and automation (\S \ref{sec:llm-automation}) settings on \ds{}.} 
\emph{Human Baseline} refers to original human-written Community Notes. 
Evaluation measures: \textbf{R = relevance, C = correctness, H = helpfulness (\S \ref{sec:hierarchical-eval})}. 
Model groups: G1 = closed-source LRMs, G2 = closed-source LLMs, G3 = open-source LLMs, G4 = domain-specific medical LLMs. 
$\dagger$ denotes reasoning-enabled models; 
\textbf{Identical R scores under Note Auto. indicate shared retriever LLM for query generation and utility judgment (see \S \ref{app:evidence-setup} and Table~\ref{tab:retriever-generator})}. 
Best and second-best results are shown in \textbf{bold} and \underline{underline}.}
\label{tab:main-results}
\end{table*}

\section{Experiments}
\label{sec:experiments}

We benchmark 15 representative LLMs against a \textbf{Human Baseline} of original health community notes. The models span four categories: \textbf{(1) closed-source large reasoning models} (LRMs) such as o3 \cite{openai2025o3}, \textbf{(2) closed-source LLMs} such as GPT-4.1 \cite{openai2025gpt4_1}, \textbf{(3) open-source LLMs and LRMs} such as Qwen3 \cite{yang2025qwen3}, and \textbf{(4) domain-specific medical LLMs} such as MedGemma \cite{sellergren2025medgemma}. We evaluate two settings: \textbf{Augmentation} (\S \ref{sec:llm-augmentation}), where models generate notes using human-provided evidence, and \textbf{Automation} (\S \ref{sec:llm-automation}), where models retrieve their own evidence. 

To ensure fair comparison in the automation setting, \textit{we restrict the retrieval quota and search timeframe to match the exact conditions available to the human note author}. Finally, to reflect platform constraints, all generated notes are strictly truncated to the 280-character limit during helpfulness evaluation. Detailed model specifications, evidence retrieval configurations, and constraint setups are provided in \textbf{Appendix \ref{app:exp-setup}}.

\paragraph{Main Results. }
Table~\ref{tab:main-results} summarizes performance across both generation modes. We highlight six observations.  
\textbf{(1)} Models perform substantially worse on the \textit{Not Helpful} subset, confirming its higher difficulty. 
\textbf{(2)} Human-written notes rated 100\% \textit{Helpful} by the crowd achieve only 73.19\% under our framework, revealing weaknesses in current voting (see \S\ref{sec:eval-superiority} for further analysis).  
\textbf{(3)} Models with over 14B parameters surpass humans in helpfulness, demonstrating the effectiveness of both augmentation and automation (see details in \S\ref{sec:gen-effectiveness}).  
\textbf{(4)} For closed-source LRMs and LLMs, automation consistently outperforms augmentation, suggesting that with well-guided retrieval, models can independently compose grounded notes.  
\textbf{(5)} The reasoning-enabled o3 model achieves highest overall scores, indicating benefits from explicit reasoning traces.  
\textbf{(6)} Domain-specific models such as MedGemma-27B outperform general-purpose models (e.g., Qwen3-32B), especially on \textit{Not Helpful} cases, reflecting stronger medical grounding.

\section{Discussion}
Building on the comparative performance results in \S\ref{sec:experiments}, we now turn to a deeper analysis of our framework's components. We structure this discussion around three key research questions (RQs):

\begin{itemize}
    \item \textbf{RQ1: Evaluation Reliability (\S \ref{sec:eval-superiority}):} How does \ours{} identify and address validity gaps in crowd ratings via hierarchical evaluation?
    \item \textbf{RQ2: Generation Quality (\S \ref{sec:gen-effectiveness}):} To what extent does \ours{} improve note correctness and helpfulness?
    \item \textbf{RQ3: Evidence Utility (\S \ref{sec:evidence-utility}): } How does the utility of evidence retrieved by \ours{} compare to human-provided sources? 
\end{itemize}

\begin{figure}[t]
    \centering
    \includegraphics[width=\columnwidth]{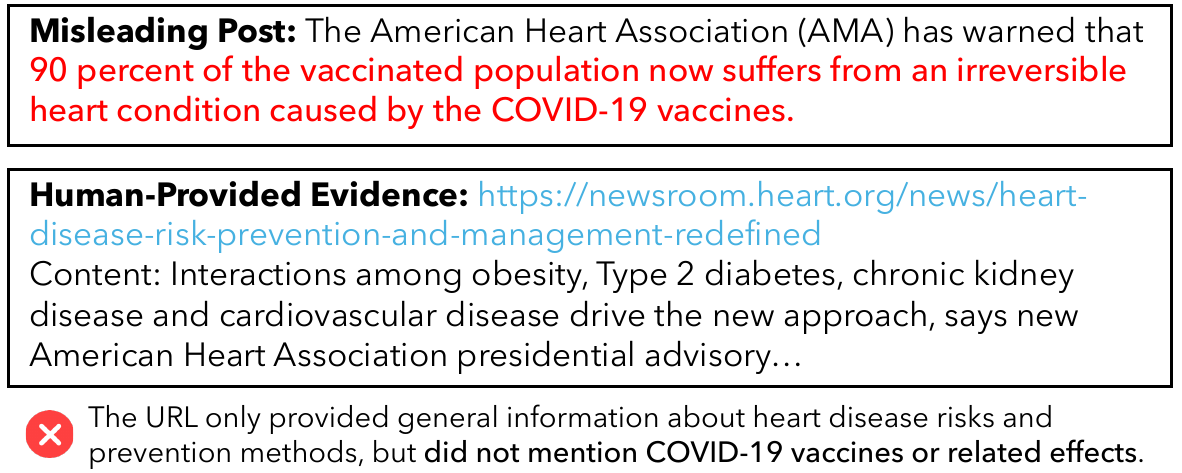}
\caption{\textbf{Example of a human-written note mislabeled as \textit{Helpful} by human voters} but correctly identified as \textit{Not Helpful} by \ours{} due to citing irrelevant evidence.}
    \label{fig:relevance-err}
\end{figure}
\begin{figure}[t]
    \centering
    \includegraphics[width=\columnwidth]{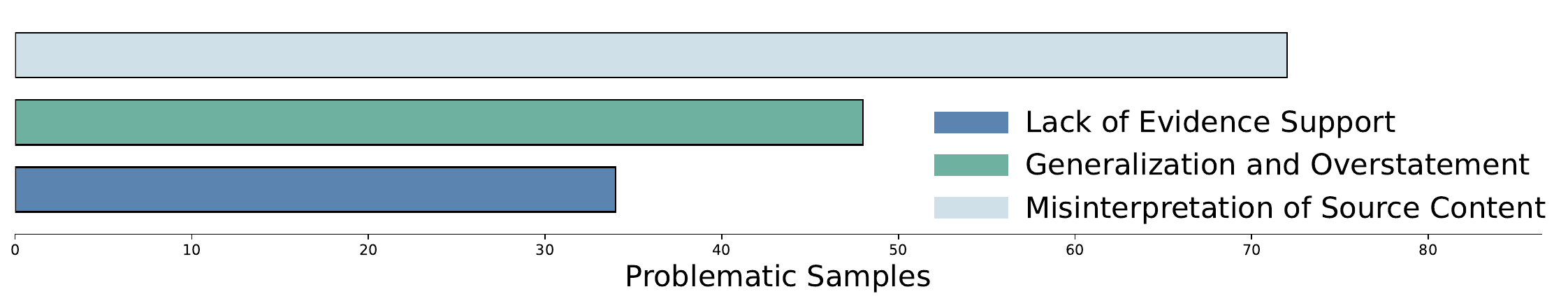}

\caption{\textbf{Error distribution of 89 human-written notes that misrepresented evidence}, grouped by three primary causes.}
    \label{fig:human-errors-dist}
\end{figure}

\subsection{\ours{} Addresses Loopholes in Crowd-Sourced Helpfulness Evaluation}
\label{sec:eval-superiority}

Our hierarchical evaluation (\S\ref{sec:hierarchical-eval}) reveals a key limitation in Community Notes voting: \textbf{many notes rated as \textit{Helpful} fail basic relevance or correctness.} As shown in Table~\ref{tab:main-results}, our framework aligns closely with human judgments on \textit{Not Helpful} (5.5\% divergence), but drops substantially on \textit{Helpful}: 11.7\% for relevance and 14.0\% for correctness.

To investigate these inconsistencies, we analyze two types of failures among notes mislabeled by humans as ``\textit{Helpful}.'' First, some notes exhibit little to no meaningful connection between their claims and the cited evidence (Figure~\ref{fig:relevance-err}), indicating weak or spurious grounding in supporting sources. Second, we conduct a focused qualitative analysis of 89 notes that our framework deems relevant but incorrect, yet were judged helpful by humans, to better understand systematic errors and common annotation pitfalls. Two human experts independently reviewed these cases and reached consensus on error attribution. As shown in Figure~\ref{fig:human-errors-dist}, three recurring causes emerge: \textbf{(1)} \textit{Lack of Evidence Support}, where claims are not substantiated by the cited sources; \textbf{(2)} \textit{Misinterpretation of Source Content}, where factual details are distorted or selectively presented; and \textbf{(3)} \textit{Overgeneralization}, where notes draw conclusions not warranted by the evidence.

These findings suggest that human voters often reward stylistic fluency over factual rigor when judging helpfulness, potentially leading to overestimation of note quality. By enforcing staged checks for relevance and correctness before assessing helpfulness, \ours{} mitigates this bias, substantially reduces false positives, and provides a more reliable and interpretable basis for helpfulness evaluation.

\subsection{\ours{} Produces Better Notes}
\label{sec:gen-effectiveness}

We next evaluate \ours{} in two settings: \textbf{(1)} augmentation (\S\ref{sec:llm-augmentation}), where models write notes from human-provided evidence, and \textbf{(2)} automation (\S\ref{sec:llm-automation}), where models retrieve evidence and generate notes end to end.

\begin{figure}[t]
    \centering
    \includegraphics[width=\columnwidth]{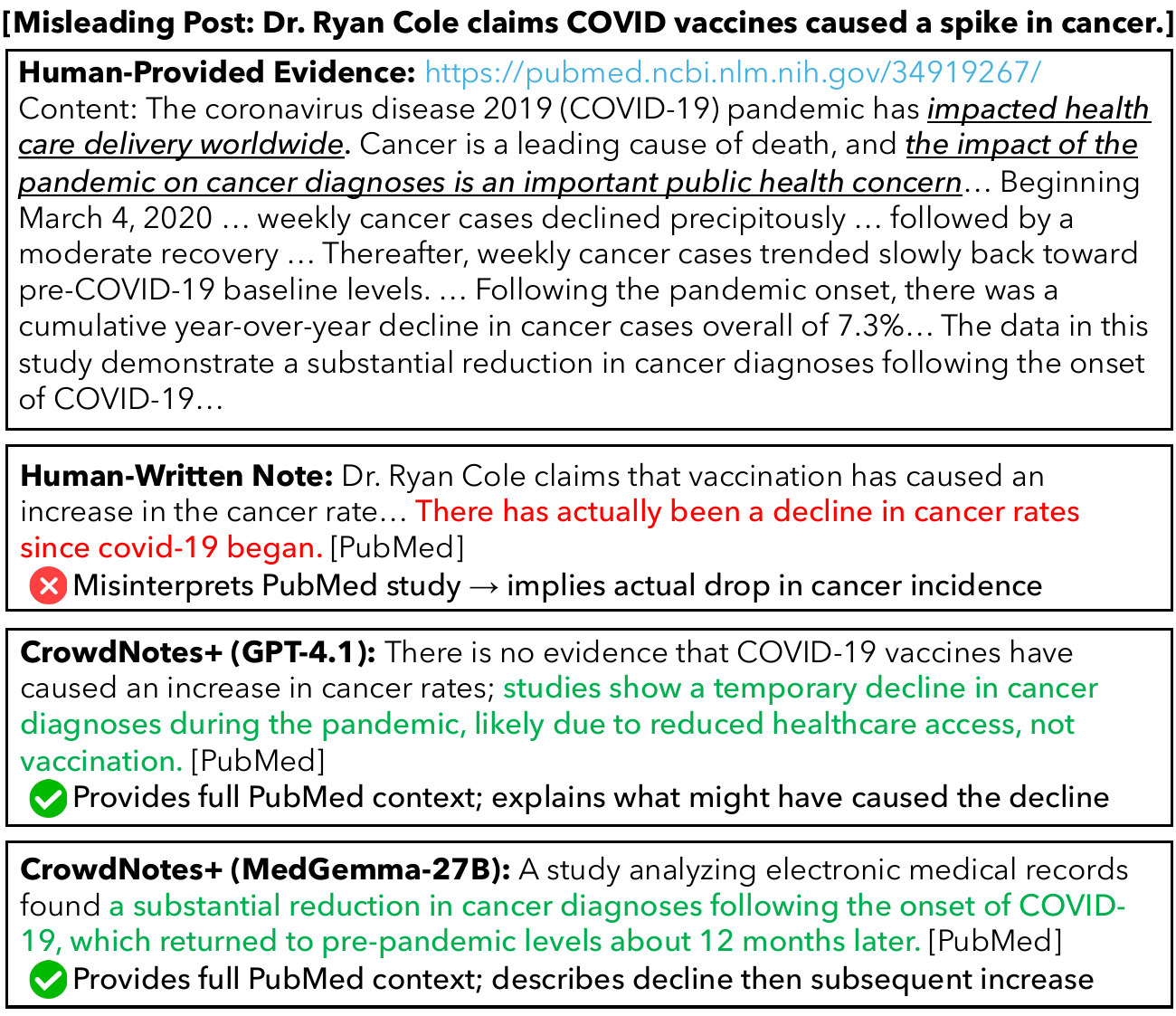}
\caption{\textbf{Effectiveness of \ours{} augmentation:} Given the same evidence, the note generated by \ours{} supplies complete contextual information that the human-written note omits.}

    \label{fig:llm-aug-benefits}
\end{figure}

\paragraph{Better Use of the Same Evidence.}
Under augmentation, \ours{} produces more correct notes than humans given the same evidence (Table~\ref{tab:main-results}), indicating stronger faithfulness to sources and clearer use of context. Figure~\ref{fig:llm-aug-benefits} shows that \ours{} often recovers key details omitted in human-written notes, improving completeness and interpretability.

\begin{table}[t]
\centering
\resizebox{\columnwidth}{!}{%
\begin{tabular}{lcc|c}
\toprule
\textbf{Model} & \textbf{Helpful} & \textbf{Not Helpful} & \textbf{Overall} \\
\midrule
\ours{} (o3) & \textbf{92.11} & \textbf{70.19} & \textbf{81.15} \\
\quad- Query Diversity & 79.50 & 69.09 & 74.30 \\
\qquad- Utility Judgment & 79.02 & 64.83 & 71.93\\
\midrule
\ours{} (MedGemma-27B) & \textbf{79.81} & \textbf{58.68} & \textbf{69.25} \\
\quad- Query Diversity & 74.76 & 54.73 & 64.75\\
\qquad- Utility Judgment & 66.25 & 50.47 & 58.36\\
\bottomrule
\end{tabular}
}
\caption{\textbf{Effectiveness of \ours{} automation:} Ablation performance in note helpfulness (\%) of utility-guided note automation in \ours{}.}
\label{tab:ablation}
\end{table}

\paragraph{Query Diversity and Utility Judgment Both Matter.}
In automation, both retrieval components are important. As shown in Table~\ref{tab:ablation}, removing either query diversity or utility judgment significantly reduces helpfulness. Query diversity expands the evidence pool, while utility judgment prioritizes the most informative and reliable sources.

\paragraph{Humans Prefer \ours{} Notes.}
We also conduct a human evaluation in automation mode (\S\ref{sec:llm-automation}), comparing human-written notes with notes generated by \ours{}. Three annotators performed pairwise comparisons on 100 randomized and anonymized note pairs from \ds{}, including 50 \textit{Helpful} and 50 \textit{Not Helpful} cases. For each pair, they selected the better note based on accuracy, relevance, specificity, neutrality, and helpfulness. Table~\ref{tab:human_note_preference} reports win rates computed from the aggregated annotator votes. The results align with our automatic evaluation: annotators consistently prefer notes generated by \ours{}, indicating higher-quality explanatory context than human-written notes.

\begin{table}[t]
\centering
\small
\begin{tabular}{lc}
\toprule
\textbf{Model} & \textbf{Win Rate (\%)} \\
\midrule
\ours{} (o3) & 87.0 \\
\ours{} (GPT-4.1) & 77.0 \\
\ours{} (MedGemma-27B) & 62.0\\
\bottomrule
\end{tabular}
\caption{\textbf{Human preference for \ours{} notes on 100 note pairs.} Win rate denotes the percentage of pairwise comparisons in which annotators preferred \ours{} over the human-written note.}
\label{tab:human_note_preference}
\end{table}

\begin{figure}[t]
    \centering
    \includegraphics[width=\columnwidth]{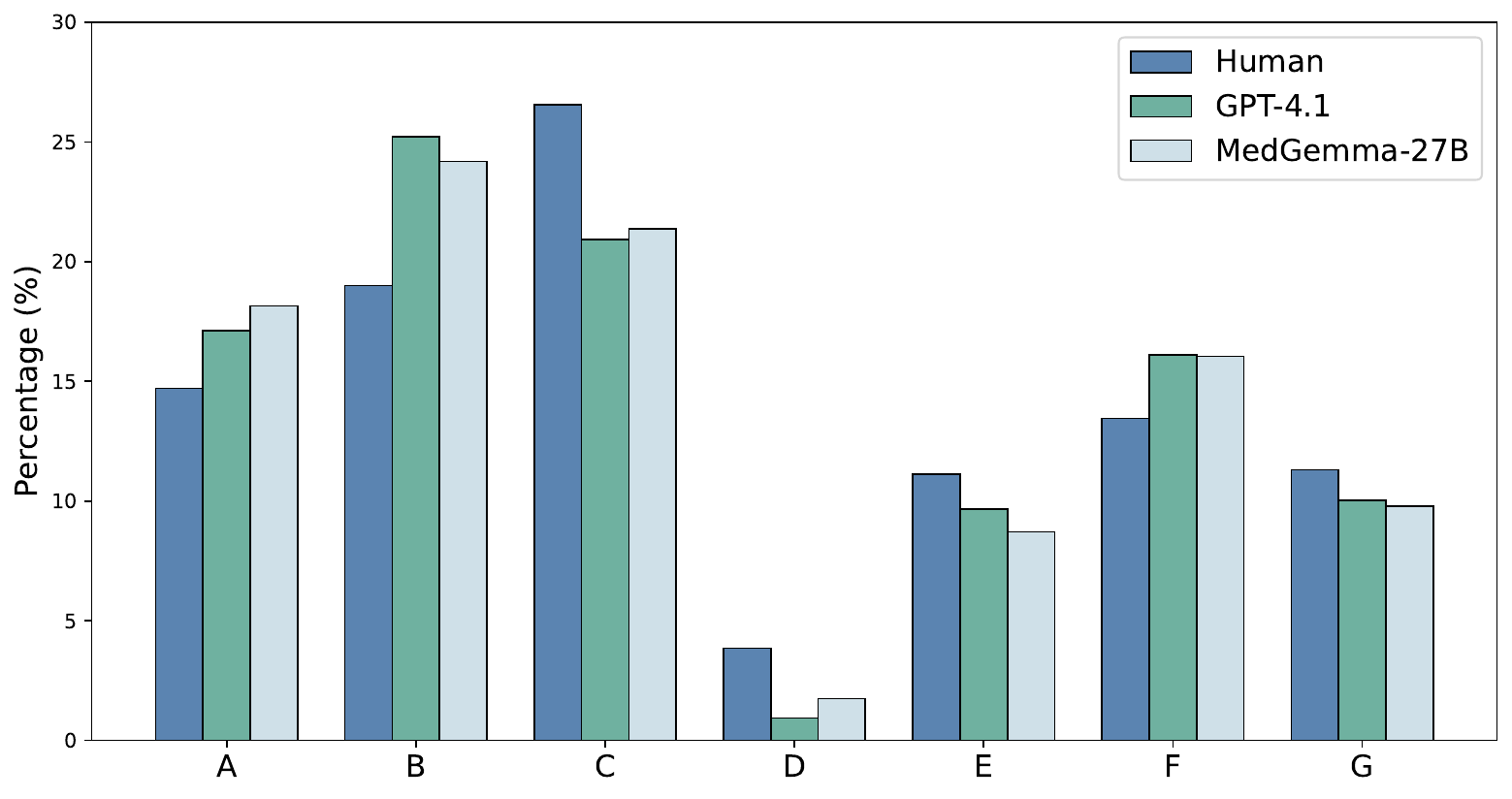}
\caption{\textbf{Comparison of human-selected and \ours{} evidence sources.} A: Health Authorities; B: Research Literature; C: News Media; D: Social Media; E: Health Portals; F: Commercial / Advocacy / NGO Sites; G: Others.}
    \label{fig:evidence-domain-dist}
\end{figure}

\subsection{\ours{} Selects Better Evidence}
\label{sec:evidence-utility}

In order to understand whether \ours{} retrieves better supporting evidence than average human contributors, we compare human-selected and \ours{}-selected evidence along two dimensions: \textbf{(1)} source distribution and \textbf{(2)} practical utility. We first examine where the evidence comes from, analyzing the types and credibility of sources used, then assess how useful it is for producing helpful, well-grounded notes across diverse misinformation scenarios and contexts.

\paragraph{\ours{} Locates More Authoritative Sources.}
We compare human- and \ours{}-selected evidence across seven categories: \textbf{(1)} Health Authorities, \textbf{(2)} Research Literature, \textbf{(3)} News Media, \textbf{(4)} Social Media, \textbf{(5)} Health Portals, \textbf{(6)} Commercial / Advocacy / NGO Sites, and \textbf{(7)} Others. A web-enabled GPT-4.1 assigns each source to a primary category. As shown in Figure~\ref{fig:evidence-domain-dist}, humans rely more on news, social media, and general health portals, while LLMs favor institutional and agency sources. This shift toward more authoritative evidence helps explain why automation consistently outperforms augmentation (Table~\ref{tab:main-results}).

\begin{table}[t]
\centering
\small
\resizebox{\columnwidth}{!}{%

\begin{tabular}{lccc}
\toprule
 \textbf{Model (vs. Human)} & \textbf{Win} & \textbf{Lose} & \textbf{Tie} \\
\midrule
\ours{} (o3)  & \textbf{65.85} & 22.48 & 11.67 \\
\ours{} (MedGemma-27B) & \textbf{57.57} & 33.20 & 9.23 \\
\bottomrule
\end{tabular}}
\caption{\textbf{Overall, \ours{} selects higher-utility evidence than humans}, demonstrated through pairwise comparisons (\%) of evidence utility between human-provided and LLM-selected sources (Figure \ref{fig:overview}).}
\label{tab:source-win-rate}

\end{table}

\begin{figure}[t]
    \centering
    \includegraphics[width=\columnwidth]{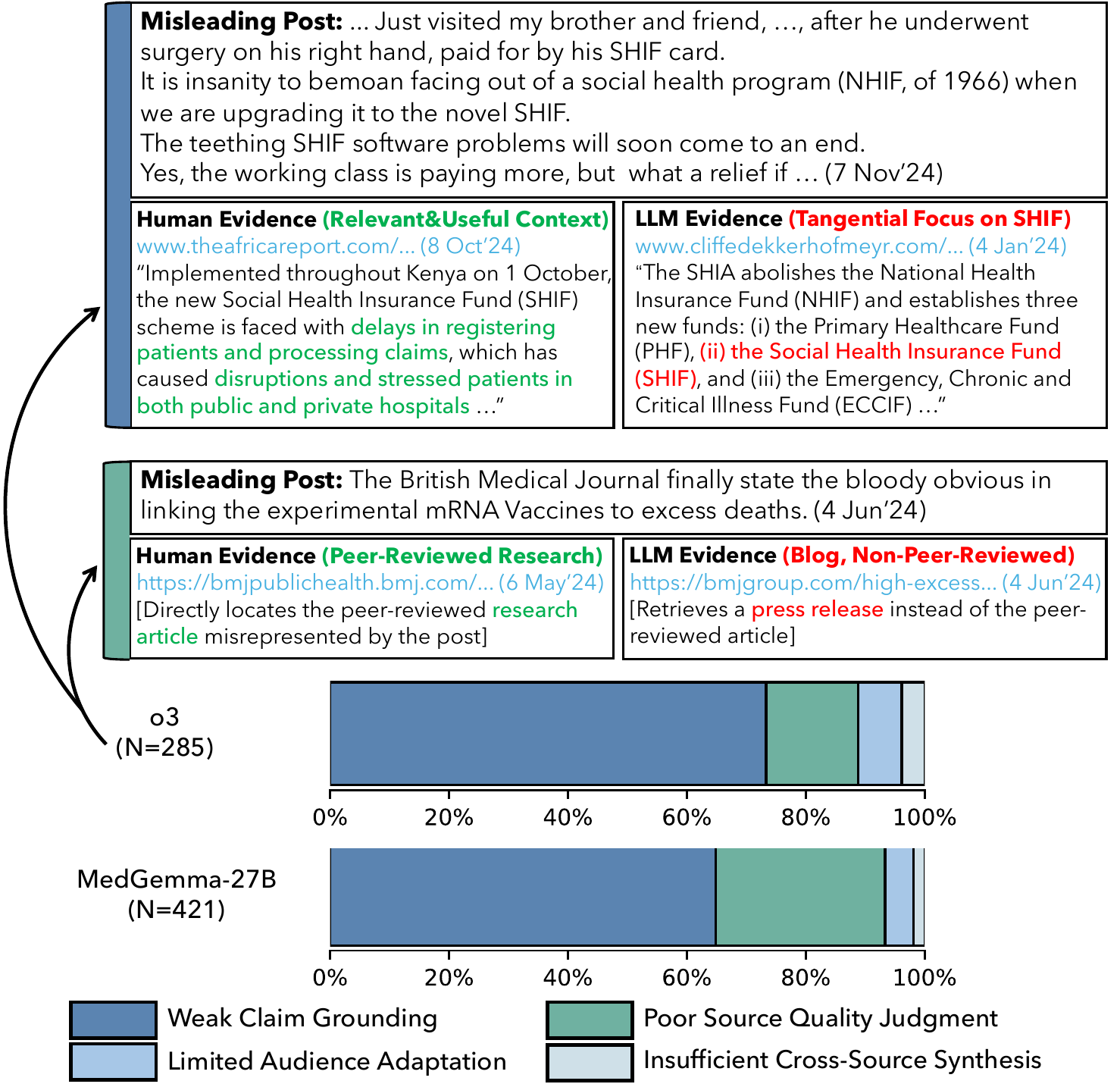}
\caption{\textbf{Why human-provided evidence is sometimes preferred over evidence selected by \ours{}.} Distribution of failure types among cases where human-provided sources are judged more useful.}
    \label{fig:llm-source-limit}
\end{figure}

\paragraph{\ours{} Often Outperforms Human Evidence Selection.}
To quantify utility, we compare human $\mathcal{E}_h$ and machine-selected evidence $\mathcal{E}_m$ on 1,268 samples in \ds{}. A web-enabled GPT-4.1 judge evaluates which set better supports helpful notes, with \ours{} using o3 and MedGemma-27B. Table~\ref{tab:source-win-rate} shows win rates above 50\%, i.e.,~\ours{} often matches or surpasses human evidence selection.

We next examine cases where human evidence is preferred. Two human experts reviewed 100 such cases and identified four recurring causes: \textbf{(1)} \textit{Weak Claim Grounding}, where the LLM misses the core claim or retrieves loosely relevant evidence; \textbf{(2)} \textit{Poor Source Quality Judgment}, where the model fails to distinguish strong from weak sources; \textbf{(3)} \textit{Limited Audience Adaptation}, where sources are overly technical or inaccessible; and \textbf{(4)} \textit{Insufficient Cross-Source Synthesis}, where multiple sources are not integrated into a coherent conclusion. The remaining cases were labeled using GPT-4.1.

Figure~\ref{fig:llm-source-limit} highlights the first two causes with examples. In the \textbf{Weak Claim Grounding} case, a post praises Kenya’s transition from NHIF to SHIF based on anecdotal experience. Human evidence directly addresses this by citing reporting on delays in SHIF registration and claims processing that disrupted services, whereas o3 retrieves a high-level overview with tangential relevance. In the \textbf{Poor Source Quality Judgment} case, the post misrepresents a study as linking mRNA vaccines to excess deaths. Humans retrieve the original peer-reviewed BMJ article, while the LLM selects a secondary press release, reflecting weaker source judgment.

Overall, while LLMs often select high-utility evidence (as shown in Table \ref{tab:source-win-rate}), their remaining failures point to shallow retrieval or insufficient integration across evidence sources, suggesting room for improvement in query formulation, multi-hop reasoning, and credibility-aware search.

\vspace{-0.2em}
\section{Conclusion and Future Work}
\vspace{-0.2em}

We identify a substantial latency gap in crowd-sourced health Community Notes, where a median delay of 17.6 hours causes corrective interventions to trail misinformation spread. To address this, we propose \ours{}, a unified framework that augments note generation and evaluation through evidence grounding, utility-guided automation, and hierarchical assessment. Experiments on \ds{} show that \ours{} can produce more accurate and helpful notes than human contributors, while also exposing a key weakness in current crowd evaluation, where note fluency is often mistaken for factual accuracy. These findings support a shift toward human--AI collaboration (see \textbf{Appendix~\ref{app:discussions}}), in which LLMs act as evidence-grounded assistants that improve the speed and reliability of community-based moderation, with clear paths for extension across domains, languages, and integrated detection pipelines.

\clearpage
\section*{Limitations}

Our work offers an important first step toward LLM-augmented Community Notes in the health domain, pointing to several extensions that could broaden its scope and practical impact. \textit{First, our investigation focuses on health content in the English language.} While health misinformation provides a high-stakes and relatively well-defined setting, applying \ours{} to more subjective domains (e.g., political or socio-cultural discourse) or to low-resource languages may introduce additional challenges related to ambiguity, cultural context, and consensus formation that are not captured in this study.

\textit{Second, although \ours{} improves evidence utility over human contributors, it remains constrained by the reasoning capabilities of current LLMs in evidence retrieval.} As observed in \S\ref{sec:evidence-utility}, models may rely on surface-level lexical overlap rather than deeper semantic understanding when selecting evidence, which can limit performance on complex or multi-hop claims. Addressing this limitation will likely require advances in retrieval models, better query formulation, and stronger integration of reasoning during evidence selection.

\textit{Finally, we evaluate \ours{} as a standalone module for advancing note creation and helpfulness assessment.} We do not model upstream detection or prioritization of misleading posts, which are critical components for real-time deployment. Integrating \ours{} into a full pipeline that includes early detection, prioritization, and intervention remains an important direction for future work.

\section*{Ethical Considerations}

\paragraph{Potential Harms and Safety.}
Although \ours{} is designed to mitigate health misinformation, deploying generative models in medical contexts carries inherent risks. A central concern is \textbf{hallucination}, where a model may produce fluent but inaccurate notes. If surfaced without oversight, such errors could lead to real-world harm. To mitigate this risk, \textit{we position \ours{} strictly as a human-augmenting system rather than a fully autonomous decision-maker.} We explicitly discourage end-to-end automation in health misinformation governance and treat human verification of retrieved evidence as a required safety layer.

\paragraph{Automation Bias.}
While our study identifies the ``fluency trap'' in human voting, introducing AI assistance introduces the complementary risk of \textbf{automation bias}, where moderators may over-trust model outputs due to their authoritative tone. Rapid generation may also incentivize speed over careful scrutiny. To counteract this risk, \textit{future interfaces built on \ours{} should promote active human engagement}, for example by requiring moderators to inspect or validate specific evidence snippets rather than simply approving generated notes.

\paragraph{Dual Use and Fairness.}
Automated fact-checking technologies have inherent dual-use potential. The same retrieval and generation mechanisms could be misused to produce persuasive, citation-backed disinformation or to selectively suppress legitimate scientific debate through biased evidence selection. In addition, reliance on indexed English-language sources may introduce \textbf{western-centric bias}, potentially under-representing non-English or local health authorities. \textit{Ongoing auditing of retrieval sources and deliberate inclusion of diverse perspectives are therefore essential.}

\paragraph{Compliance with Platform Policies.}
All data collection and usage in this work comply with platform policies and public data guidelines. X posts and web evidence were obtained through authorized APIs and exclude private or personally identifiable information. To balance reproducibility with user privacy, we will release \ds{} under controlled, research-only access.

\section*{Acknowledgments}
This research is supported by the Ministry of Education, Singapore, under its Academic Research Fund Tier 1 (T1 251RES2508) and MOE AcRF TIER 3 Grant (MOE-MOET32022-0001). We thank Sahajpreet Singh (National University of Singapore) for early conversations related to the Community Notes concept.

\bibliography{custom}

\clearpage
\appendix

\section{Discussion: Implications for Human--AI Collaborative Misinformation Governance}
\label{app:discussions}

\paragraph{LLMs as End-to-End Assistants in the Note Creation Pipeline.} 
Our findings in \S \ref{sec:eval-superiority} and \S \ref{sec:gen-effectiveness} suggest that integrating LLMs into Community Notes for \textbf{(1)} evidence selection, \textbf{(2)} note generation, and \textbf{(3)} hierarchical evaluation can \textit{substantially improve the relevance, correctness, and helpfulness of crowd-sourced misinformation mitigation.}

\paragraph{LLM Support for Evidence Selection and Note Generation.} 
As discussed in \S \ref{sec:gen-effectiveness}, the quality and appropriateness of evidence play a central role in shaping note accuracy. When LLMs are given the same human-selected sources (Figure \ref{fig:llm-aug-benefits}), they are able to organize and synthesize this evidence more effectively during note generation. Building on this foundation, the strong performance of \textbf{utility-guided automation} (\S \ref{sec:evidence-utility}, Table \ref{tab:main-results}) shows that LLMs can also enhance the evidence selection process itself by retrieving more authoritative and contextually relevant sources. These improvements in evidence availability and quality naturally lead to notes with stronger factual grounding. Future refinements such as intent-aware search \cite{wang2025beyond} and query diversification \cite{wu2024result} may further strengthen this evidence foundation and support even more reliable note generation.

\paragraph{LLM Support for More Reliable Evaluation.} Recent commentary from the X Community Notes Team \cite{li2025scaling} envisions a hybrid workflow that continues to rely on human voting for helpfulness assessment. In contrast, our analysis in \S\ref{sec:eval-superiority} shows that human voting often rewards stylistic fluency even when factual support is weak. \ours{} addresses this issue through a \textbf{hierarchical evaluation pipeline} (\S\ref{sec:hierarchical-eval}) that assesses relevance, correctness, and helpfulness in sequence using reliable judges (Appendix~\ref{app:judge-reliability}), producing more reliable and interpretable note assessments.

\paragraph{Toward Hybrid Human--AI Governance.}
Taken together, these findings point to a hybrid human--AI misinformation governance model in which LLMs provide factual rigor, high-quality evidence selection, and consistent first-pass evaluation, while human contributors contribute oversight, social context, and pluralistic perspectives. Such a division of responsibilities offers a path toward more scalable, timely, and trustworthy misinformation governance.

\section{The \ds{} Benchmark}
\label{app:healthnotes}

Using the 1,268 human-written health-related notes described in \S \ref{sec:benchmark}, we leverage their corresponding post IDs from the public Community Notes dataset\footnote{\url{https://x.com/i/communitynotes/download-data}}
 to retrieve the associated flagged posts via the X API.

\begin{figure}[t]
    \centering
    \includegraphics[width=\columnwidth]{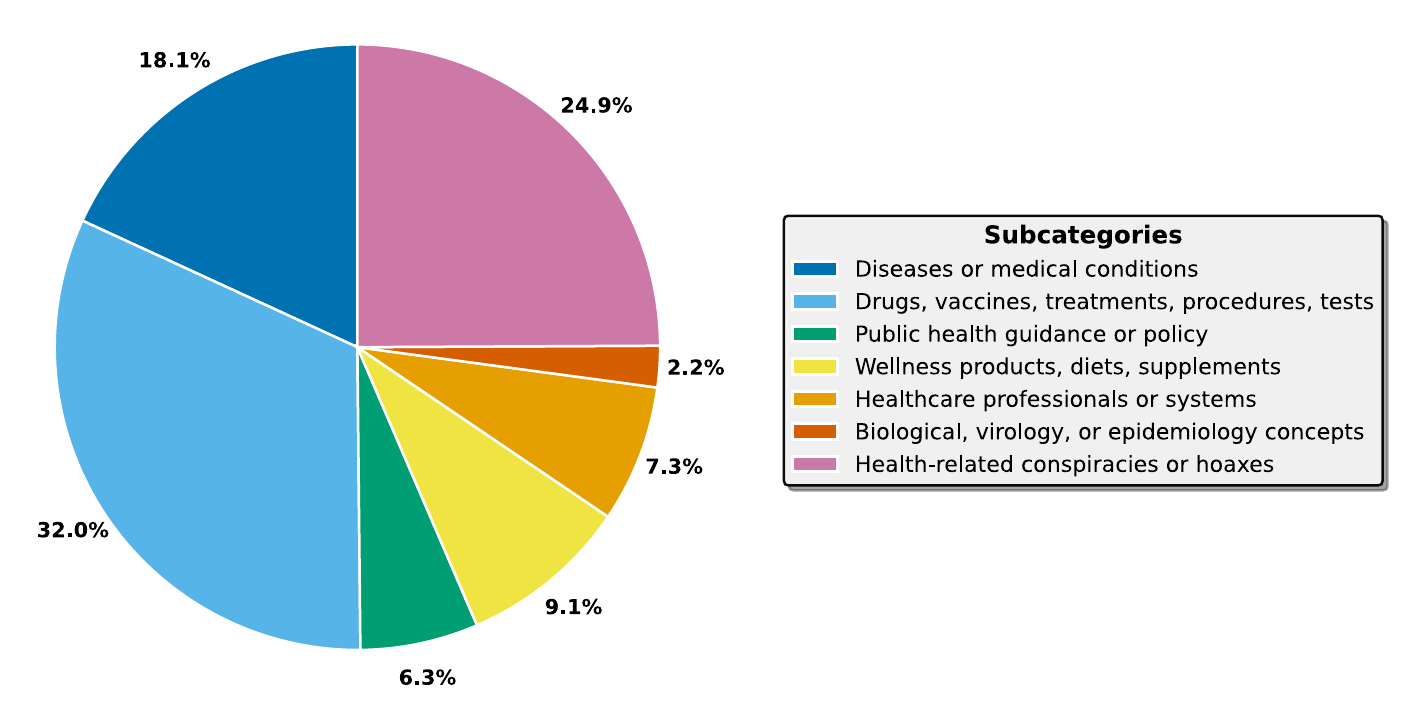}
    \caption{Topic distribution of notes in \ds{}.
}  
    \label{fig:topic-dist}
\end{figure}

Table~\ref{tab:data-stats} summarizes core statistics of \ds{}. To examine topical coverage, we group notes by the primary category assigned during the filtering step (following the seven major health-related categories defined in \S \ref{sec:data-scope}). As shown in Figure~\ref{fig:topic-dist}, \ds{} spans a broad range of medical and public health issues. Three categories—diseases or medical conditions, public health guidance and policy, and health-related conspiracies or hoaxes—are particularly prominent, reflecting the types of claims that frequently generate community attention and require timely clarification on social media.
\begin{table}[t]
\centering
\small
\begin{tabular}{lccc}
\toprule
 & \textbf{\#. of Notes} & \textbf{\#. of Posts} & \textbf{\#. of URLs} \\
\midrule
\textbf{Helpful} & 634 & 608 & 1,330 \\
\textbf{Not Helpful} & 634 & 622 & 907 \\
\bottomrule
\end{tabular}
\caption{Dataset statistics for \ds{}. Notes span May 2022–Aug 2025, and their corresponding posts span Jun 2020–Jul 2025.}
\label{tab:data-stats}
\end{table}

\section{Details of Note Generation in \ours{}}
\label{app:note-gen-detail}

This section provides additional details on how \ours{} constructs notes in both augmentation and automation modes. We describe \textbf{(1)} how evidence is curated through utility-guided selection (Appendix \ref{app:evidence-acq}), \textbf{(2)} how retrieved webpages are processed into evidence chunks (Appendix \ref{app:evidence-retrieval}), and \textbf{(3)} how LLMs synthesize these chunks into contextual notes (Appendix \ref{app:note-gen}).

\subsection{Utility-Guided Evidence Curation}
\label{app:evidence-acq}

In the automation mode (\S \ref{sec:llm-automation}), evidence is sourced from the Web through a utility-guided selection process rather than human-provided URLs as in the augmentation mode (\S \ref{sec:llm-augmentation}). Given a candidate pool $\mathcal{P}$ of evidence snippets (each containing a webpage title and summary from Google Custom Search API\footnote{\url{https://developers.google.com/custom-search/}}), an LLM estimates the \textit{utility} of each snippet for supporting or contextualizing the flagged post. The prompt template used for utility judgment is shown below:
\begin{prompt}{}
\small  
SYSTEM\_PROMPT = ``You are a careful selector. Output exactly ONE integer as instructed.'' \\

You are selecting one source (healthcare-related Community Note utility).

This is selection round \#\{round\_no\}. Choose exactly ONE result that has the highest utility. \\

\#\# Utility should reflect whether the search result is:

- Relevant to the tweet's topic.

- Likely to add meaningful background or clarification.

- Reliable enough to be worth retrieving. \\

\#\# OUTPUT FORMAT (critical):

- Output EXACTLY one integer, the index of your chosen item (1..\{len(items\_remaining)\}).
 
- No extra words. No numbering other than the single integer. No explanations. \\

\#\# Tweet:
\{tweet\} \\

\#\# Search Results (candidates):

[\{idx\}] Title: \{title\}
        Snippet: \{snippet\}
        URL: \{url\}

\end{prompt}

Across $\tau$ iterative rounds, the highest-utility snippet is selected and removed from $\mathcal{P}$, yielding a final quota of $\tau$ evidence items. The URLs associated with these items form the machine-selected evidence set $\mathcal{E}_m$, which is subsequently used for retrieval and note generation. The distributional differences between human- and LLM-selected evidence are shown in Figure~\ref{fig:evidence-domain-dist}.
\subsection{Evidence Retrieval and Processing}
\label{app:evidence-retrieval}

For each evidence set, whether human-provided ($\mathcal{E}_h$) or LLM-selected ($\mathcal{E}_m$), we retrieve the corresponding webpages using the Jina API\footnote{\url{https://jina.ai/}}. Retrieved pages are cleaned to remove non-essential elements such as headers, footers, navigation bars, and reference sections. The remaining body text is segmented into overlapping passages of 512 tokens with a 128-token overlap.

Each 512-token passage is embedded using \texttt{sentence-transformers/all-mpnet-base-v2}, and the most semantically similar passage to the flagged post $p$ is selected per source. These form the evidence chunks $\mathcal{C}_h$ (human) or $\mathcal{C}_m$ (LLM), used for note generation.

\subsection{Note Generation}
\label{app:note-gen}

Given the evidence chunks, either human-provided ($\mathcal{C}_h$) or LLM-retrieved ($\mathcal{C}_m$), \ours{} generates contextual notes for flagged posts identified as potentially misleading. Both the augmentation and automation settings (\S \ref{sec:llm-augmentation} and \S \ref{sec:llm-automation}) employ the same prompt template for note generation:
\begin{prompt}{}
\small  
SYSTEM\_PROMPT = ``Community notes is a collaborative way to add helpful context to posts and keep people better informed. Now you are a highly experienced community note writer.'' \\

Task: Write a community note based ONLY on the source snippets below.

Hard constraints:

- The note MUST be in English.

- DO NOT include any URLs in the note.

- The note MUST be a single line (no line breaks, no bullets).

- Note length MUST be $\leq$ \{budget\_chars\} characters. Do not exceed this budget.

- Be specific, objective, and verifiable. \\

Tweet:
\{tweet\} \\

Source snippets:

[S\{index\}] {url} (chunk \{chunk\_id\})
\{text\}

Output only the note content. Remember: length $\leq$ \{budget\_chars\}, no URLs.

\end{prompt}

The model conditions on the flagged post $p$ and the selected evidence chunks to produce a concise, fact-grounded explanation. The generated note text is paired with its corresponding evidence URLs in the final output, ensuring transparency and traceability in line with Community Notes conventions.

\section{Details of Hierarchical Evaluation in \ours{}}
\label{app:eval-detail}

As introduced in \S \ref{sec:hierarchical-eval}, \ours{} uses a three-step hierarchical evaluation where a note advances only after passing the previous stage. This appendix details the stages: \textbf{(1)} evidence relevance (Appendix \ref{app:eval-relevance}), \textbf{(2)} evidence representation correctness (Appendix \ref{app:eval-correctness}), and \textbf{(3)} note helpfulness (Appendix \ref{app:eval-helpfulness}). We also report \textbf{human and automated evaluations validating the reliability of the judge models} (Appendix \ref{app:judge-reliability}).

\subsection{Evidence Relevance}
\label{app:eval-relevance}

\paragraph{Setup.} The relevance stage assesses whether the retrieved evidence provides meaningful factual context or clarification that helps readers evaluate the claim made in the post. We use GPT-4.1 to perform this assessment via the following prompt:
\begin{prompt}{}
\small  
SYSTEM\_PROMPT = ``You are a very meticulous inspector.''

You are given a Tweet and one or more Source snippets:
Tweet:
\{tweet\} \\

Source snippets:

[S\{index\}] \{url\} (chunk \{chunk
\_id\})

\{text\} \\

Task: Determine whether any of the Source snippets adds meaningful factual background, clarification, or supporting information that helps better understand or evaluate the claim made in the Tweet.

1. Check each snippet independently.

2. If at least one snippet meets the requirements, output ``Final decision: yes''; otherwise output ``Final decision: no''.

\end{prompt}

\subsection{Evidence Representation Correctness}
\label{app:eval-correctness}

Conditioned on passing relevance, we next evaluate whether the note accurately represents the cited sources, avoiding factual errors, exaggeration, and misleading framing. This step also uses GPT-4.1 with the prompt shown below:

\begin{prompt}{}
\small  

SYSTEM\_PROMPT = "You are a very meticulous inspector" \\

You are given a Community note and one or more Source snippets:

Community Note:

\{note\}

Source snippets:

[S\{index\}] \{url\} (chunk \{chunk\_id\})
\{text\} \\

Task: Decide whether the Community note distorts the information in any of the provided Source snippets.

1. Check each snippet independently.

2. If at least one distortion is found, output ``Final decision: yes''; otherwise output ``Final decision: no''.

\end{prompt}

\subsection{Note Helpfulness}
\label{app:eval-helpfulness}

Conditioned on passing correctness, the final stage assesses whether a note provides useful context that helps readers understand or critically evaluate the flagged post, following the official Community Notes guidelines. We use \judge{} (a fine-tuned Lingshu-7B model \cite{li2025scaling}) with temperature 0 for deterministic and domain-adapted scoring. 

To mirror platform constraints, \textbf{this is the only stage where the 280-character limit used by Community Notes is applied}: if the note text and URLs exceed 280 characters (each URL counts as one), the text is truncated before evaluation.

\begin{prompt}{}
\small  
SYSTEM\_PROMPT = ``You are a precise text classifier.'' \\ 

You are given a Tweet and its corresponding Note:

Tweet: \{tweet\}  

Note: \{note (only note text, w/o evidence URLs)\} \\

The purpose of note is to add helpful context to tweet and keep people better informed.
Your task is to evaluate whether the Note is Helpful or Not Helpful based on the following criteria: \\

**Helpful Criteria:** 

- Clear and/or well-written 

- Cites high-quality sources

- Directly addresses the Tweet's claim 

- Provides important context 

- Neutral or unbiased language

- Other (any additional positive reason)  \\

**Not Helpful Criteria:**

- Incorrect information

- Sources missing or unreliable

- Misses key points or irrelevant

- Hard to understand

- Argumentative or biased language

- Spam, harassment, or abuse

- Sources do not support note

- Opinion or speculation

- Note not needed on this Tweet

- Other (any additional negative reason) \\

Instructions:

1. Carefully read the Tweet and the Note.

2. Analyze the Note using the Helpful and Not Helpful criteria above.

3. Respond with ``Final decision: yes'' (if Helpful) or ``Final decision: no'' (if Not Helpful).

\end{prompt}

\paragraph{\judge{} Training Setup} 
\judge{} is trained on human-labeled health-related post–note pairs, using only note text (without appended URLs) to ensure that helpfulness judgments reflect explanatory quality rather than evidence relevance or correctness. The dataset contains 2,971 Helpful and 742 Not Helpful post–note pairs, with 1,000 pairs (800 Helpful, 200 Not Helpful) reserved for evaluation.

Each instance is formatted as a chat prompt using the helpfulness evaluation template, with loss applied only to the final decision tokens (``Final decision: yes/no'') and left padding for causal alignment. Training uses full fine-tuning for 2 epochs with AdamW (learning rate $1\times 10^{-5}$), gradient accumulation of 16 steps, and bfloat16 precision.

The resulting model produces deterministic, parseable outputs suitable for automatic evaluation and consistent downstream analysis. \textit{Although some posts in \ds{} overlap with those present in \judge{}'s training data, all associated notes in \ds{} are distinct, ensuring that no helpfulness labels or note content leak into evaluation.} This separation prevents label leakage and preserves a fair assessment of generalization to unseen note formulations.

\begin{table}[t]
\centering
\resizebox{\columnwidth}{!}{%
\begin{tabular}{lcc}
\toprule
\textbf{Model} & \textbf{Macro-F1 (\%)} & \textbf{Macro-Accuracy (\%)} \\
\midrule
GPT-4.1 & 74.28 & 74.19 \\
Gemini-2.5-flash & 68.36 & 65.13 \\
Claude-Sonnet-4 & 78.14 & 76.44 \\
Lingshu-32B & 64.71 & 62.25 \\
Lingshu-7B & 51.66 & 51.63 \\
\midrule
\judge{}& \textbf{81.03} & \textbf{81.44} \\
\bottomrule
\end{tabular}
}
\caption{\textbf{Effectiveness of \judge{} for note helpfulness assessment}, validated by its superior performance on 1,000 unseen post–note pairs.}
\label{tab:judge-perf}
\end{table}

\subsection{Judge Reliability Assessment}
\label{app:judge-reliability}

This section evaluates the reliability of the judge models used at each stage of the hierarchical evaluation in \ours{}. For relevance and correctness, we assess LLM-as-a-Judge decisions through human evaluation to verify consistency with expert judgments. For helpfulness, we measure \judge{}'s alignment with human-labeled ground truth, providing a quantitative assessment of its accuracy and robustness in capturing human notions of note quality.

\subsubsection{Reliability of Relevance Judgments} 
\label{app:reliability-relevance}

In order to assess the reliability of the LLM-based evidence relevance judgments (see Appendix \ref{app:eval-relevance} for details), we conduct a human evaluation on 100 sampled model predictions: 50 notes from the \textit{Helpful} subset of \ds{} and 50 from the \textit{Not Helpful} subset (see \S \ref{sec:benchmark}), ensuring balanced coverage of both positive and negative cases. This sampling strategy allows us to evaluate model behavior across a diverse range of relevance scenarios.

Three graduate student annotators independently labeled each instance following standardized instructions designed to ensure consistency and minimize ambiguity across judgments. The annotation protocol, including detailed criteria and examples, is provided as follows.

\begin{prompt}{}
\small  
\textbf{Human Evaluation Objective.} The goal of this evaluation is to assess whether the reasoning produced by the LLM judge provides a reasonable and sufficient justification for its final predicted \textbf{relevance label}. \\

Each data instance contains the following fields: 
\begin{itemize}[leftmargin=*]
    \item id: data identifier.
    \item tweet: the text of the flagged post.
    \item evidence\_snippets: retrieved evidence snippets (each with a URL and its associated text chunk)
    \item relevance\_label: the LLM’s predicted relevance label. ``Yes'' indicates at least one evidence snippet is relevant to the tweet, and ``No'' indicates all evidence snippets are irrelevant.
    \item reasoning: the LLM’s explanation supporting its prediction.
\end{itemize}

\textbf{For reference, the exact prompt used for LLM judgment is reproduced below:}

\{prompt for evaluating evidence relevance, presented in Appendix \ref{app:eval-relevance}\} \\

\textbf{Annotation Guidelines.} For each instance, you will need to assign one of two labels:
\begin{itemize}[leftmargin=*]
    \item \textbf{0 (Reliable):} (a) The reasoning is coherent and clearly articulated. (b) The relevance label is consistent with the reasoning. (c) The final decision is acceptable to a human annotator.
    \item \textbf{1 (Unreliable):} (a) The reasoning is inconsistent with the evidence or the tweet. (b) The reasoning does not justify the final relevance label. (c) There are clear logical errors, misinterpretations, or unsupported conclusions in the LLM reasoning trace.
\end{itemize}

\end{prompt}

We report the \textbf{agreement rate} between the LLM judge and majority human annotations. \textbf{The LLM matches the aggregated judgment in all 100 cases.} Inter-annotator disagreement occurs in only one instance (majority \textit{Reliable}). As a verification task where high agreement is expected, this serves as a sanity check of the LLM judge’s consistency with human assessments.

\subsubsection{Reliability of Correctness Judgments}  
\label{app:reliability-correctness}

To evaluate the reliability of LLM-based correctness judgments (Appendix \ref{app:eval-correctness}), we follow a similar procedure as Appendix \ref{app:reliability-relevance}. We sample 100 correctness judgments made by the model: 50 notes derived from posts in the \textit{Helpful} subset and 50 notes from the \textit{Not Helpful} subset.

The same three annotators from Appendix \ref{app:reliability-relevance} independently assessed whether the LLM's justification and decision accurately reflected the provided sources, using the following instructions.

\begin{prompt}{}
\small  
\textbf{Human Evaluation Objective.} The goal of this evaluation is to determine whether the reasoning produced by the LLM provides a reasonable and sufficient justification for its final predicted \textbf{distortion label}. \\

Each data instance contains the following fields: 
\begin{itemize}[leftmargin=*]
    \item id: data identifier.
    \item note: the Community Note text.
    \item evidence\_snippets: retrieved evidence snippets (each with a URL and its associated text chunk)
    \item distortion\_label: the LLM’s prediction of whether the note distorts the evidence. ``Yes'' indicates that the note contains at least one instance of misrepresenting the evidence, and ``No'' indicates that the note does not distort any provided evidence.
    \item reasoning: the LLM’s explanation supporting its prediction.
\end{itemize}

\textbf{For reference, the exact prompt used for LLM judgment is reproduced below:}

\{prompt for evaluating evidence representation correctness, presented in Appendix \ref{app:eval-correctness}\} \\

\textbf{Annotation Guidelines.} For each instance, you will need to assign one of two labels:
\begin{itemize}[leftmargin=*]
    \item \textbf{0 (Reliable):} (a) The reasoning is coherent and clearly articulated. (b) The relevance label is consistent with the reasoning. (c) The final decision is acceptable to a human annotator.
    \item \textbf{1 (Unreliable):} (a) The reasoning conflicts with the content of the note or the evidence. (b) The reasoning does not justify the final distortion label. (c) There are clear logical errors, misinterpretations, or unsupported conclusions in the LLM reasoning trace.
\end{itemize}

\end{prompt}

As with the relevance evaluation, we report the \textbf{agreement rate} between the LLM judge and majority-voted human annotations as the primary reliability metric. \textbf{The LLM prediction matches the aggregated human judgment in 99 out of 100 cases, indicating strong alignment.} Inter-annotator disagreement occurs in only 3 cases, comprising 2 instances with a majority \textit{Reliable} label and 1 instance with a majority \textit{Unreliable} label. Given that this is a verification task where high agreement is expected, these results serve as a sanity check, confirming the LLM judge’s consistency and robustness relative to human assessments.

\subsubsection{Reliability of Helpfulness Judgments} 
\label{app:reliability-helpfulness}
For the final stage, we evaluate \judge{} by comparing its \textit{Helpful}/\textit{Not Helpful} predictions with human-contributed labels on the 1,000 test samples described in Appendix \ref{app:eval-helpfulness}. 

As shown in Table~\ref{tab:judge-perf}, \judge{} achieves higher alignment with human judgments than GPT-4.1, Claude-4-Sonnet \cite{anthropic2025claude4}, and Gemini 2.5 Flash \cite{google2025gemini_2_5}. These results demonstrate strong reliability for domain-specific helpfulness evaluation.

\section{Experimental Setup}
\label{app:exp-setup}

This section details the setup for evidence acquisition (Appendix \ref{app:evidence-setup}), note generation (Appendix \ref{app:models-evaluated}), and evaluation constraints (Appendix \ref{app:eval-constraints}) used in \ours{} experiments.

\subsection{Evidence Acquisition Setup}
\label{app:evidence-setup}

For the \textbf{Automation} setting described in \S \ref{sec:llm-automation}, we select six representative LLMs to perform utility-guided evidence retrieval: o3, GPT-4.1, Qwen3 (32B and 8B), and MedGemma (27B and 4B). Correlations between Retriever LLMs and Generator LLMs are summarized in Table \ref{tab:retriever-generator}.

To ensure fair comparison with human-provided evidence, we apply the following controls:
\begin{itemize}[leftmargin=*]
\item \textbf{Quota Matching:} The evidence quota $\tau$ for each sample equals the number of URLs in the human evidence set ($|\mathcal{E}_h|$).
\item \textbf{Temporal Restrictions:} \textit{Web search results are constrained to content available \textbf{up to the timestamp of the human-written note}}, preventing access to future information.
\item \textbf{Passage Extraction:} For each retrieved webpage, we extract the highest-ranked 512-token passage to serve as the evidence snippet for synthesizing notes. 
\end{itemize}

\begin{table}[t]
\centering
\small
\begin{tabular}{l|c}
\toprule
\textbf{Evidence Retriever} & \textbf{Note Generator} \\
\midrule
\multirow{3}{*}{o3 $\dagger$} 
  & Gemini-2.5-pro $\dagger$  \\
  & o3 $\dagger$ \\
  & Grok-4 $\dagger$\\
\midrule
\multirow{2}{*}{GPT-4.1} 
  & GPT-4.1 \\
  & Claude-4-Opus \\
\midrule
\multirow{2}{*}{Qwen3-32B} 
  & Qwen3-32B \\
  & Qwen3-14B \\
\midrule
\multirow{4}{*}{Qwen3-8B} 
  & Llama-3.1-8B \\
  & Ministral-8B \\
  & Qwen3-8B $\dagger$ \\
  & Qwen3-8B \\
\midrule
\multirow{2}{*}{MedGemma-27B} 
  & Lingshu-32B \\
  & MedGemma-27B \\
\midrule
\multirow{2}{*}{MedGemma-4B} 
  & Lingshu-7B \\
  & MedGemma-4B \\
\bottomrule
\end{tabular}
\caption{\textbf{Correlation between retriever LLMs (used for query generation and utility judgment) and generator LLMs (used for note generation) in the \textbf{Automation} setting.} This mapping explains identical relevance scores observed across certain generator models (see Table~\ref{tab:main-results}). $\dagger$ denotes reasoning-enabled models.}
\label{tab:retriever-generator}
\end{table}

\subsection{Note Generation Setup} 
\label{app:models-evaluated}
Under both \textbf{Augmentation} (\S \ref{sec:llm-augmentation}) and \textbf{Automation} (\S \ref{sec:llm-automation}) settings, we evaluate \textbf{15 representative LLMs} grouped into four categories ([G1] to [G4]):
\begin{itemize}[leftmargin=*] 
    \item \textbf{[G1] Closed-Source Large Reasoning Models (LRMs):} Models trained with chain-of-thought or extensive reasoning capabilities, including o3~\cite{openai2025o3}, Gemini-2.5~\cite{google2025gemini_2_5}, and Grok-4~\cite{xai2025grok4}. 
    \item \textbf{[G2] Closed-Source LLMs:} Standard state-of-the-art proprietary models, specifically GPT-4.1~\cite{openai2025gpt4_1} and Claude-4~\cite{anthropic2025claude4}. 
    \item \textbf{[G3] Open-Source LLMs and LRMs:} High-performing open weights models, including Qwen3~\cite{yang2025qwen3}, Llama-3.1~\cite{dubey2024llama}, and Ministral~\cite{mistral2024ministral}.
    \item \textbf{[G4] Domain-Specific Medical LLMs:} Models fine-tuned for biomedical contexts, such as Lingshu~\cite{xu2025lingshu} and MedGemma~\cite{sellergren2025medgemma}. 
\end{itemize} 

Unless otherwise specified, we use non-reasoning variants of open-source models with temperature set to 0 to ensure deterministic and reproducible outputs, and run all experiments once under this setting. Detailed model specifications, including parameter sizes and configurations, are listed in Table~\ref{tab:model_cards}.

\begin{table}[t]
\centering
\small
\resizebox{\columnwidth}{!}{%
\begin{tabular}{l|c}
\toprule
\textbf{Model} & \textbf{Model Card} \\
\midrule
Gemini-2.5-Pro & \texttt{gemini-2.5-pro-preview-03-25} \\
o3 & \texttt{o3-2025-04-16} \\
Grok-4 & \texttt{x-ai/grok-4} \\
\midrule
GPT-4.1 & \texttt{gpt-4.1-2025-04-14} \\
Claude-Opus-4 & \texttt{claude-opus-4-20250514} \\
\midrule
Qwen3-32B & \texttt{Qwen/Qwen3-32B}\\
Qwen3-14B & \texttt{Qwen/Qwen3-14B} \\
Llama-3.1-8B & \texttt{meta-llama/Llama-3.1-8B-Instruct} \\
Ministral-8B & \texttt{mistralai/Ministral-8B-Instruct-2410} \\
Qwen3-8B & \texttt{Qwen/Qwen3-8B}  \\
\midrule
Lingshu-32B & \texttt{lingshu-medical-mllm/Lingshu-32B} \\
MedGemma-27B & \texttt{
google/medgemma-27b-text-it} \\
Lingshu-7B & \texttt{lingshu-medical-mllm/Lingshu-7B} \\
MedGemma-4B & \texttt{
google/medgemma-4b-it} \\
\bottomrule
\end{tabular}
}
\caption{The model versions of LLMs used in \ours{} for note generation.}
\label{tab:model_cards}
\end{table}

\subsection{Note Length Constraints} 
\label{app:eval-constraints}

Community Notes imposes a strict character limit of 280. We mirror this in our evaluation: 
\begin{itemize}[leftmargin=*] 
\item \textbf{Constraint Application:} If the combined length of an LLM-generated note and its appended URLs exceeds 280 characters, we truncate the text content. Following X's policy, URLs count as a single character\footnote{\url{https://docs.x.com/x-api/community-notes/quickstart}}. 
\item \textbf{Evaluation Scope:} This truncation applies only to the \textit{Helpfulness} evaluation. We do not truncate notes for \textit{Relevance} or \textit{Correctness} evaluations, as these metrics assess the logical validity of the generated content rather than its final presentation format. 
\end{itemize}

\section{Demonstrations of \ours{} Workflow}
\label{app:ours-examples}

\begin{figure*}[t]
    \centering
    \includegraphics[width=0.8\textwidth]{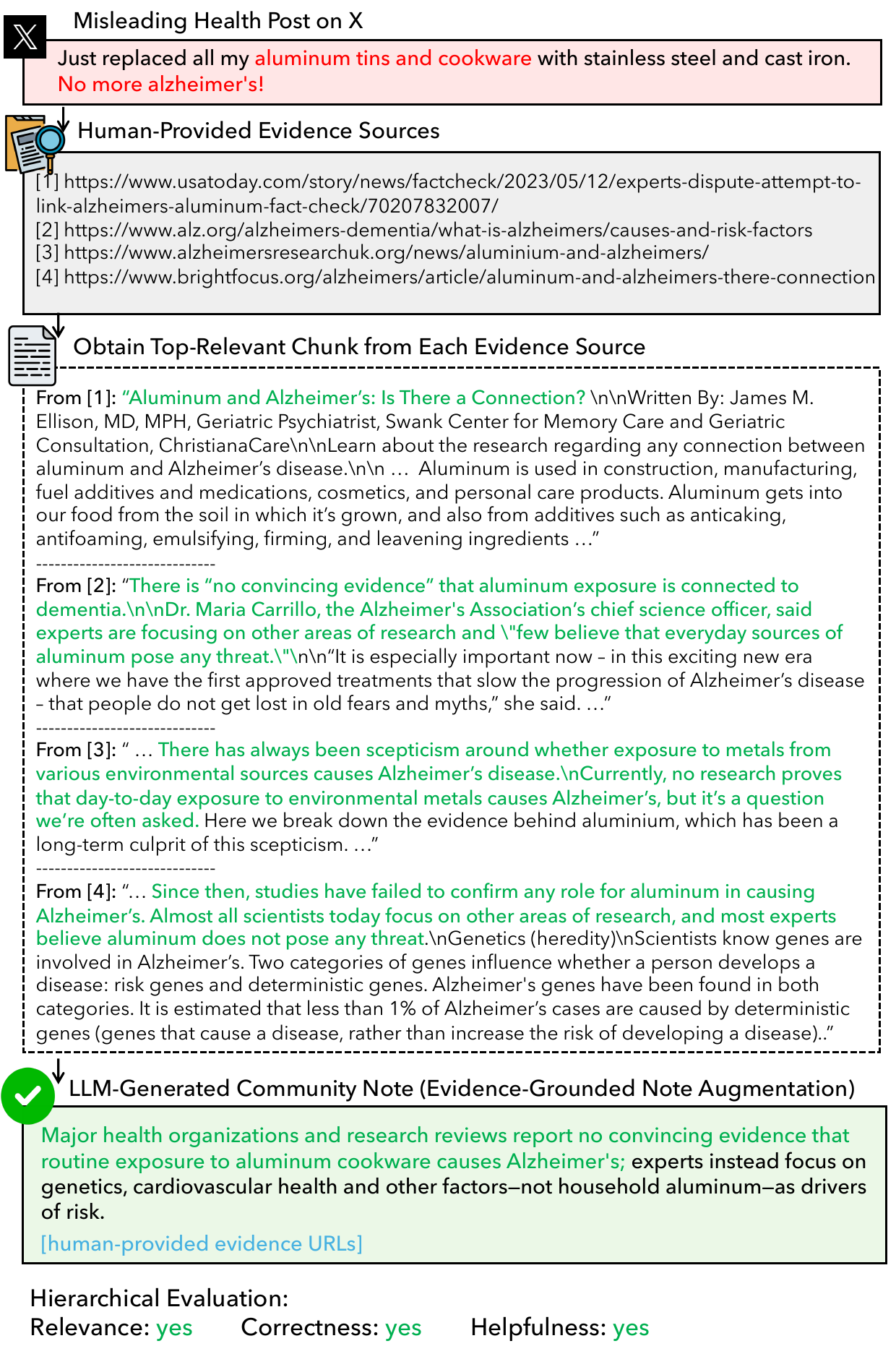}
\caption{\textbf{Illustration of \ours{} under the evidence-grounded augmentation setting (\S \ref{sec:llm-augmentation}).} Using evidence chunks retrieved from human-provided sources, the o3 model synthesizes the information to generate a helpful note, which addresses the post's misleading claim that aluminum exposure causes Alzheimer’s disease.}
\label{fig:aug-example}
\end{figure*}

\begin{figure*}[t]
    \centering
    \includegraphics[width=0.8\textwidth]{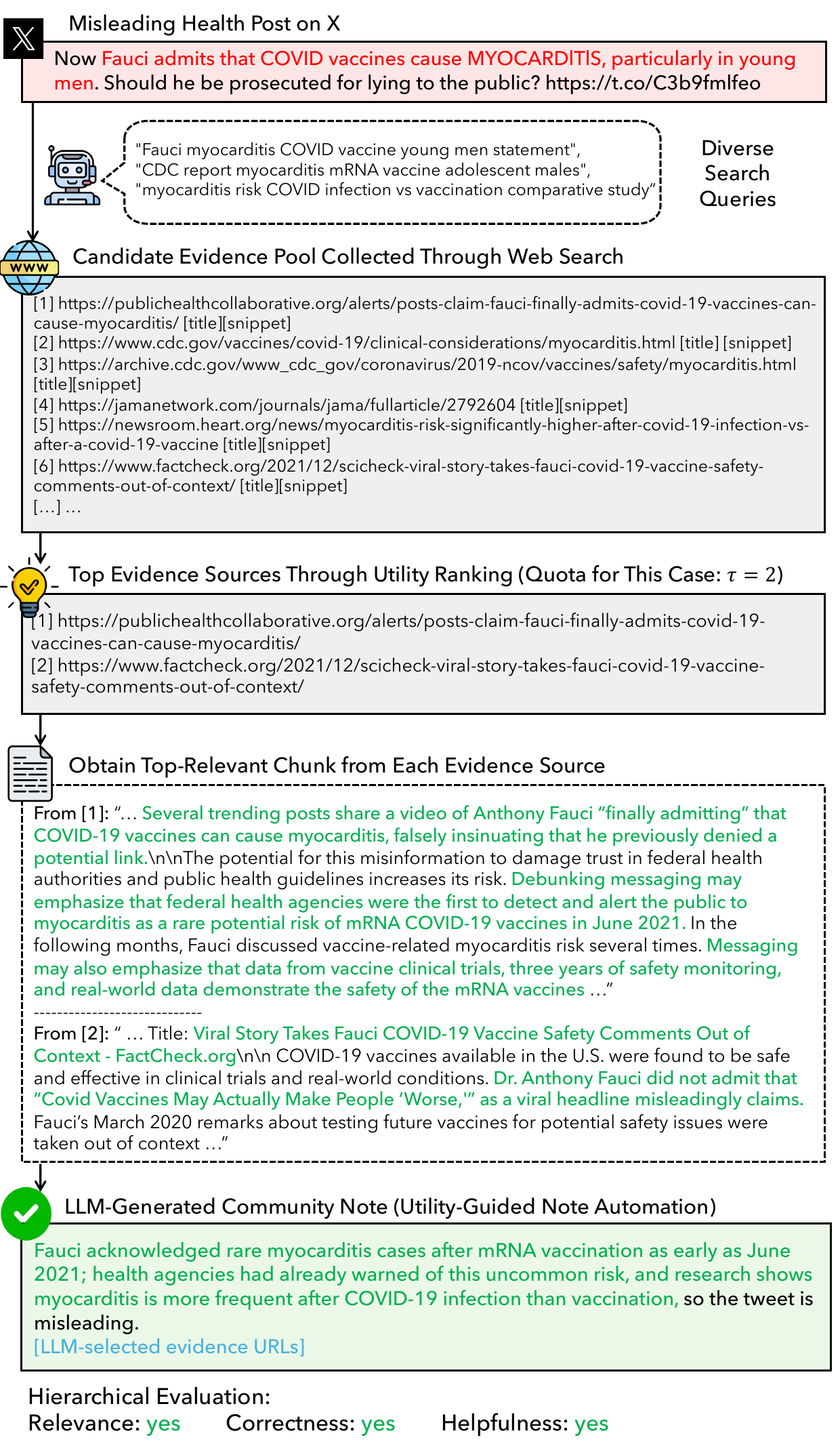}
\caption{\textbf{Illustration of \ours{} under the utility-guided automation setting (\S \ref{sec:llm-automation}).} Using evidence chunks retrieved from LLM-selected sources, the o3 model synthesizes the information to generate a helpful note addressing the misleading claim that Fauci ``admitted'' COVID vaccines cause myocarditis. \textbf{For a fair comparison with human-written notes, the evidence quota for this case is set to $\tau = 2$} to match the number of human-provided sources.}
\label{fig:auto-example}
\end{figure*}

We present two end-to-end examples that illustrate how \ours{} performs evidence acquisition, note generation, and hierarchical evaluation. Figure \ref{fig:aug-example} illustrates a case from our evidence-grounded note augmentation setting (\S \ref{sec:llm-augmentation}), and Figure \ref{fig:auto-example} illustrates a case from our utility-guided note automation setting (\S \ref{sec:llm-automation}). 

\clearpage

\end{document}